\documentstyle[12pt,epsf]{article}

  \def\pp{{\mathchoice
          {
              \kern 1pt%
              \raise 1pt
              \vbox{\hrule width5pt height0.4pt depth0pt
                    \kern -2pt
                    \hbox{\kern 2.3pt
                          \vrule width0.4pt height6pt depth0pt
                          }
                    \kern -2pt
                    \hrule width5pt height0.4pt depth0pt}%
                    \kern 1pt
           }
            {
              \kern 1pt%
              \raise 1pt
              \vbox{\hrule width4.3pt height0.4pt depth0pt
                    \kern -1.8pt
                    \hbox{\kern 1.95pt
                          \vrule width0.4pt height5.4pt depth0pt
                          }
                    \kern -1.8pt
                    \hrule width4.3pt height0.4pt depth0pt}%
                    \kern 1pt
            }
            {
              \kern 0.5pt%
              \raise 1pt
              \vbox{\hrule width4.0pt height0.3pt depth0pt
                    \kern -1.9pt  
                    \hbox{\kern 1.85pt
                          \vrule width0.3pt height5.7pt depth0pt
                          }
                    \kern -1.9pt
                    \hrule width4.0pt height0.3pt depth0pt}%
                    \kern 0.5pt
            }
            {
              \kern 0.5pt%
              \raise 1pt
              \vbox{\hrule width3.6pt height0.3pt depth0pt
                    \kern -1.5pt
                    \hbox{\kern 1.65pt
                          \vrule width0.3pt height4.5pt depth0pt
                          }
                    \kern -1.5pt
                    \hrule width3.6pt height0.3pt depth0pt}%
                    \kern 0.5pt
            }
        }}

  \def\mm{{\mathchoice
                       {
                             \kern 1pt
               \raise 1pt    \vbox{\hrule width5pt height0.4pt depth0pt
                                  \kern 2pt
                                  \hrule width5pt height0.4pt depth0pt}
                             \kern 1pt}
                       {
                            \kern 1pt
               \raise 1pt \vbox{\hrule width4.3pt height0.4pt depth0pt
                                  \kern 1.8pt
                                  \hrule width4.3pt height0.4pt depth0pt}
                             \kern 1pt}
                       {
                            \kern 0.5pt
               \raise 1pt
                            \vbox{\hrule width4.0pt height0.3pt depth0pt
                                  \kern 1.9pt
                                  \hrule width4.0pt height0.3pt depth0pt}
                            \kern 1pt}
                       {
                           \kern 0.5pt
             \raise 1pt  \vbox{\hrule width3.6pt height0.3pt depth0pt
                                  \kern 1.5pt
                                  \hrule width3.6pt height0.3pt depth0pt}
                           \kern 0.5pt}
                       }}

\catcode`@=11
\def\un#1{\relax\ifmmode\@@underline#1\else
        $\@@underline{\hbox{#1}}$\relax\fi}
\catcode`@=12

\let\du=\du                     

\def\a{\alpha}
\def\b{\beta}

\def\d{\delta}
\def\e{\epsilon}
\def\f{\phi}
\def\g{\gamma}
\def\h{\eta}

\def\j{\psi}
\def\k{\kappa}
\def\l{\lambda}
\def\m{\mu}
\def\n{\nu}
\def\o{\omega}
\def\p{\pi}
\def\q{\theta}
\def\r{\rho}
\def\s{\sigma}
\def\t{\tau}

\def\x{\xi}
\def\z{\zeta}
\def\D{\Delta}
\def\F{\Phi}
\def\G{\Gamma}

\def\L{\Lambda}
\def\O{\Omega}
\def\P{\Pi}

\def\S{\Sigma}

\def\ve{\varepsilon}
\def\vf{\varphi}

\def\vq{\vartheta}

\def\ca{{\cal A}}

\def\cd{{\cal D}}

\def\cf{{\cal F}}

\def\ch{{\cal H}}

\def\ck{{\cal K}}
\def\cl{{\cal L}}
\def\cm{{\cal M}}

\def\ct{{\cal T}}

\def\cv{{\cal V}}

\def\bo{{\raise-.3ex\hbox{\large$\Box$}}}               
\def\pa{\partial}                                       
\def\de{\nabla}                                         
\def\pr{\prod}                                          
\def\TH{{\raise.2ex\hbox{$\displaystyle \bigodot$}\mskip-4.7mu \llap H \;}}
\def\face{{\raise.2ex\hbox{$\displaystyle \bigodot$}\mskip-2.2mu \llap {$\ddot
        \smile$}}}                                      
\def\dg{\sp\dagger}                                     

\def\sp#1{{}^{#1}}                              
\def\Bar#1{\overline{#1}}                       
\def\VEV#1{\left\langle #1\right\rangle}        
\def\abs#1{\left| #1\right|}                    
\def\leftrightarrowfill{$\mathsurround=0pt \mathord\leftarrow \mkern-6mu
        \cleaders\hbox{$\mkern-2mu \mathord- \mkern-2mu$}\hfill
        \mkern-6mu \mathord\rightarrow$}
\def\dvec#1{\vbox{\ialign{##\crcr
        \leftrightarrowfill\crcr\noalign{\kern-1pt\nointerlineskip}
        $\hfil\displaystyle{#1}\hfil$\crcr}}}           
\def\dt#1{{\buildrel {\hbox{\LARGE .}} \over {#1}}}     

\def\frac#1#2{{\textstyle{#1\over\vphantom2\smash{\raise.20ex
        \hbox{$\scriptstyle{#2}$}}}}}                   
\def\sfrac#1#2{{\vphantom1\smash{\lower.5ex\hbox{\small$#1$}}\over
        \vphantom1\smash{\raise.4ex\hbox{\small$#2$}}}} 
\def\bfrac#1#2{{\vphantom1\smash{\lower.5ex\hbox{$#1$}}\over
        \vphantom1\smash{\raise.3ex\hbox{$#2$}}}}       
\def\afrac#1#2{{\vphantom1\smash{\lower.5ex\hbox{$#1$}}\over#2}}    

\def\[{\lfloor{\hskip 0.35pt}\!\!\!\lceil}
\def\]{\rfloor{\hskip 0.35pt}\!\!\!\rceil}

\def\du#1#2{_{#1}{}^{#2}}
\def\ud#1#2{^{#1}{}_{#2}}

\def\ha{{\fracmm12}}
\def\tr{{\rm tr}}
\def\Tr{{\rm Tr}}

\def\un{\underline}
\def\fracmm#1#2{{{#1}\over{#2}}}

\def\low#1{{\raise -3pt\hbox{${\hskip 0.75pt}\!_{#1}$}}}

\def\Dot#1{\buildrel{_{_{\hskip 0.01in}\bullet}}\over{#1}}
\def\dt#1{\Dot{#1}}

\newskip\humongous \humongous=0pt plus 1000pt minus 1000pt
\def\caja{\mathsurround=0pt}
\def\eqalign#1{\,\vcenter{\openup2\jot \caja
        \ialign{\strut \hfil$\displaystyle{##}$&$
        \displaystyle{{}##}$\hfil\crcr#1\crcr}}\,}
\newif\ifdtup

\def\sbar#1{\stackrel{*}{\Bar{#1}}}             

\def\ref#1{$\sp{#1)}$}

\def\pl#1#2#3{Phys.~Lett.~{\bf {#1}B} (19{#2}) #3}
\def\np#1#2#3{Nucl.~Phys.~{\bf B{#1}} (19{#2}) #3}
\def\prl#1#2#3{Phys.~Rev.~Lett.~{\bf #1} (19{#2}) #3}
\def\pr#1#2#3{Phys.~Rev.~{\bf D{#1}} (19{#2}) #3}
\def\cqg#1#2#3{Class.~and Quantum Grav.~{\bf {#1}} (19{#2}) #3}
\def\cmp#1#2#3{Commun.~Math.~Phys.~{\bf {#1}} (19{#2}) #3}

\def\mpl#1#2#3{Mod.~Phys.~Lett.~{\bf A{#1}} (19{#2}) #3}

\def\ibid#1#2#3{{\it ibid.}~{\bf {#1}} (19{#2}) #3}

\parskip=\medskipamount                 
\thicklines                         

\begin{document}


\thispagestyle{empty}               

\def\border{                                            
        \setlength{\unitlength}{1mm}
        \newcount\xco
        \newcount\yco
        \xco=-24
        \yco=12
        \begin{picture}(140,0)
        \put(-20,11){\tiny Institut f\"ur Theoretische Physik Universit\"at
Hannover~~ Institut f\"ur Theoretische Physik Universit\"at Hannover~~
Institut f\"ur Theoretische Physik Hannover}
        \put(-20,-241.5){\tiny Institut f\"ur Theoretische Physik Universit\"at
Hannover~~ Institut f\"ur Theoretische Physik Universit\"at Hannover~~
Institut f\"ur Theoretische Physik Hannover}
        \end{picture}
        \par\vskip-8mm}

\def\headpic{                                           
        \indent
        \setlength{\unitlength}{.8mm}
        \thinlines
        \par
        \begin{picture}(29,16)
        \put(75,16){\line(1,0){4}}
        \put(80,16){\line(1,0){4}}
        \put(85,16){\line(1,0){4}}
        \put(92,16){\line(1,0){4}}

        \put(85,0){\line(1,0){4}}
        \put(89,8){\line(1,0){3}}
        \put(92,0){\line(1,0){4}}

        \put(85,0){\line(0,1){16}}
        \put(96,0){\line(0,1){16}}
        \put(92,16){\line(1,0){4}}

        \put(85,0){\line(1,0){4}}
        \put(89,8){\line(1,0){3}}
        \put(92,0){\line(1,0){4}}

        \put(85,0){\line(0,1){16}}
        \put(96,0){\line(0,1){16}}
        \put(79,0){\line(0,1){16}}
        \put(80,0){\line(0,1){16}}
        \put(89,0){\line(0,1){16}}
        \put(92,0){\line(0,1){16}}
        \put(79,16){\oval(8,32)[bl]}
        \put(80,16){\oval(8,32)[br]}

        \end{picture}
        \par\vskip-6.5mm
        \thicklines}

\border\headpic {\hbox to\hsize{
\vbox{\noindent DESY 98 -- 109  \hfill August 1998 \\
ITP--UH--18/98 \hfill hep-th/9808187 }}}

\noindent
\vskip1.3cm
\begin{center}

{\Large\bf  Dynamical generation of gauge and Higgs bosons \newline
           in N{=}2 supersymmetric non-linear sigma-models}
\footnote{Supported in part by the `Deutsche Forschungsgemeinschaft' and the
NATO grant CRG 930789}\\
\vglue.3in

Sergei V. Ketov \footnote{
On leave of absence from:
High Current Electronics Institute of the Russian Academy of Sciences,\newline
${~~~~~}$ Siberian Branch, Akademichesky~4, Tomsk 634055, Russia}

{\it Institut f\"ur Theoretische Physik, Universit\"at Hannover}\\
{\it Appelstra\ss{}e 2, 30167 Hannover, Germany}\\
{\sl ketov@itp.uni-hannover.de}
\end{center}
\vglue.2in
\begin{center}
{\Large\bf Abstract}
\end{center}

A four-dimensional N=2 supersymmetric non-linear sigma-model with the Eguchi-Hanson (ALE) 
target space and a non-vanishing central charge is rewritten to a classically equivalent and
formally renormalizable gauged `linear' sigma-model over a non-compact coset space in N=2 harmonic 
superspace by making use of an N=2 vector gauge superfield as the Lagrange multiplier. It is then
demonstrated that the N=2 vector gauge multiplet becomes dynamical after taking into account 
one-loop corrections due to quantized hypermultiplets. This implies the appearance of a composite 
gauge boson, a composite chiral spinor doublet and a composite complex Higgs particle, all defined 
as the physical states associated with the propagating N=2 vector gauge superfield. The composite 
N=2 vector multiplet is further identified with the zero modes of a superstring ending on a 
D-6-brane. Some non-perturbative phenomena, such as the gauge symmetry enhancement for coincident 
D-6-branes and the Maldacena conjecture, turn out to be closely related to our NLSM via M-theory.
Our results support a conjecture about the composite nature of superstrings ending on D-branes.

\newpage

\section{Introduction}

The idea that some of the `elementary' particles, like a photon, Higgs or W bosons, may 
be composite is known in theoretical high-energy physics for many years, while it was 
proposed as a possible solution to many different problems in quantum field theory. For 
instance, the compositeness of photons was suggested long time ago, in order to resolve 
the ultra-violet problems of Quantum Electrodynamics related to the existence of Landau 
pole and the divergence of the effective coupling at high energies \cite{lan}. If the
Higgs particles are to be interpreted as bound states, this would simply explain the 
experimental failure to observe them, since an acceptable scale for their compositeness
is certainly much larger than any available energies. The compositeness of some of 
the vector bosons mediating weak or strong interactions was also proposed to accommodate 
the phenomenologically reguired gauge group $SU(3)\times SU(2)\times U(1)$ of the 
Standard Model (SM) in the maximally extended four-dimensional N=8 supergravity 
\cite{cju}. Gauging the internal symmetry of the N=8 supergravity merely produces $SO(8)$
as the gauge group which does not contain the SM gauge group as a subgroup \cite{wnic}. 
However, since the scalar sector of the N=8 supergravity can be described by the 
non-compact non-linear sigma-model (NLSM) over the coset $E_7/SU(8)$ \cite{van}, assuming
that its auxiliary gauge fields become dynamical in quantum theory would give rise to the
gauge group $SU(8)$ which is big enough. Though the N=8 supergravity is no longer 
considered  as the unifying quantum field theory because of its apparent 
non-renormalizability, its modern successor known under the name of M-theory \cite{wm} 
does, nevertheless, have the eleven-dimensional supergravity as the low-energy effective 
action, whose dimensional reduction down to four spacetime  dimensions yields the N=8 
supergravity. Moreover, the bound states arising in a system of the BPS-type extended
classical solutions to the eleven-dimensional supergravity (known as branes) are known to
play an important role in M-theory \cite{wm}. 

The quantum field-theoretical mechanisms of dynamical generation of composite particles
are known in two or three spacetime dimensions \cite{avl,wo}.~\footnote{See also 
ref.~\cite{pol} for an introduction.} Unfortunately, little is known about the formation 
of bound states in quantized four-dimensional field theories (see, however, 
ref.~\cite{vanh}) or in M-theory (see, however, ref.~\cite{yale}).

In the present paper I investigate the possible mechanism for a generation of composite
N=2 vector multiplets in a four-dimensional N=2 supersymmetric NLSM with an ALE target 
space. The basic idea is to reformulate this classical NLSM to the renormalizable form 
given by the gauged `linear' NLSM over a non-compact coset space, and then to take into 
account the one-loop quantum corrections due to the quantized hypermultiplets comprising 
fields of both positive and negative norm ({\it cf.} ref.~\cite{vanh}). The N=2
extended supersymmetry with a non-vanishing central charge plays the important role in 
our approach: on the one side, it implies an ALE hyper-K\"ahler geometry of the NLSM 
target space and the particular form of the associated scalar potential, whereas, on 
the other side, it automatically gives rise to many divergence cancellations which, otherwise, 
could destroy a consistency of the proposed theory. The technical power of N=2 harmonic superspace 
(HSS) allows us to take advantage of having the manifest N=2 extended supersymmetry in 
quantum perturbation theory, which singificantly simplifies our calculations and makes them 
very transparent.
 
The paper is organized as follows: in sect.~2 we discuss some known general facts about 
superspace and complex geometry of supersymmetric 4d NLSM, which are going to be relevant in
the next sections. In particular, we emphasize the relation between N=2 HSS and the hyper-K\"ahler 
geometry of N=2 NLSM, and the role of isometries in making this connection explicit. In sect.~3 we 
define several classically equivalent HSS forms of the N=2 NLSM with the Eguchi-Hanson metric and a 
non-vanishing central charge, and show their relation to a larger class of N=2 NLSM with 
multicentre (Gibbons-Hawking) metrics. In sect.~4 we quantize a coset (gauged) representation of 
the Eguchi-Hanson N=2 NLSM in HSS, and demonstrate a dynamical generation of an N=2 vector 
multiplet. A relation to M-theory and brane technology for N=2 supersymmetric quantum gauge field 
theories in 4d is discussed in sect.~5. Our conclusions and possible generalizations are outlined 
in sect.~6. A brief introduction into 4d, N=2 HSS is given in  Appendix A. In Appendix B we collect
some known facts about N=2 restricted chiral superfields and Fayet-Iliopoulos (FI) terms. A brief 
account of our main results is available in the electronically published Proceedings of the 
STRINGS'98 Conference \cite{strings}.  

\section{Complex geometry of 4d NLSM and superspace}

Let $x^{\m}$, $\m=0,1,2,3$, be the coordinates of a flat four-dimensional (4d) spacetime
of signature $(+,-,-,-)$. By definition of the bosonic 4d NLSM, its real scalar fields 
$\f^a(x^{\m})$, $a=1,2,\ldots,n$, themselves are to be considered as the coordinates of 
some (internal) NLSM target space $\cm$ of real dimension $n$. The standard action of the
4d bosonic NLSM reads
$$ S_{\rm bosonic}[\f] =\fracmm{1}{2\k^2}\int d^4x\,
 g_{ab}(\f)\pa_{\m}\f^a\pa^{\m}\f^b~,\eqno(2.1)$$
where the set of functions $g_{ab}(\f)$ is called the NLSM metric. If the fields $\f^a$ are
chosen to be dimensionless, then the coupling constant $\k$ has to be of dimension of length, in 
order to make the action dimensionless (in units $\hbar=c=1$). An important consequence of this 
fact is the non-renormalizability (by index) of the quantized field theory (2.1) with any non-flat 
NLSM metric (i.e. with a curved target NLSM space $\cm$). It follows, for example, from the 
covariant background field method in application to the theory (2.1) that its one-loop on-shell
counterterm includes some terms of the fourth-order in spacetime derivatives, while the 
field-dependent coefficient functions in front of these terms are essentially given by 
the NLSM curvature tensor squared (see e.g., ref.~\cite{nlsmbook} for a review). 
Renormalizability thus requires the NLSM curvature to vanish, which just amounts to a 
flat NLSM metric. This clearly makes quantum 4d NLSM to be very different from their 2d 
renormalizable (in some generalised sense) counterparts whose coupling constant is 
dimensionless and whose one-loop one-shell counterterm in 2d is governed by the Ricci 
tensor not the curvature \cite{nlsmbook}. A supersymmetrization of 4d NLSM (see below)
does not remove the non-renormalizability in 4d \cite{spe}. 

The action (2.1) is invariant under arbitrary (differentiable and invertible) field 
reparametrizations provided that $g_{ab}(\f)$ transforms as a second-rank tensor. 
However, no conserved Noether current is associated with this symmetry, since the 
induced transformation of $g_{ab}(\f)$ as the function of $\f$ is generically different 
from the tensor transformation law. The only exception arises when a field diffeomorphism
$\d\f=\x$ is an isometry of $\cm$, i.e. when the Lie derivative of metric vanishes, 
$\cl_{\x}g_{ab}=0$.

The N=1 supersymmetrization of the theory (2.1) is straightforward in superspace
\cite{zum}. Since the NLSM scalar fields are to belong to scalar N=1 supermultiplets
which are described by chiral N=1 superfields $\F$, $\bar{D}_{\dt{\a}}\F=0$, the NLSM 
geometry has to be complex. Moreover, on dimensional reasons, the most general NLSM 
action of the second-order in spacetime derivatives (in components) has to be governed 
in superspace by a real function of chiral superfields $\F$ and their 
conjugates (anti-chiral supefields $\bar{\F}$), i.e.
$$ S[\F,\Bar{\F}] =\fracmm{1}{\k^2}\int d^4x d^2\q d^2\bar{\q}\,K(\F,\Bar{\F})~,\eqno(2.2)$$
where the superspace measure is now of dimension two (in units of length), the coupling
constant $\k$ is still of dimension one, and all chiral superfields are dimensionless.

After rewriting eq.~(2.2) in components one finds a purely bosonic contribution of the 
form 
$$ S_{\rm bosonic}[\f,\bar{\f}] =\fracmm{1}{\k^2}\int d^4x\,
 g_{i\bar{j}} (\f,\bar{\f})\pa_{\m}\f^i\pa^{\m}\bar{\f}^{\bar{j}}~,\eqno(2.3)$$
where $\f^i$ are the leading complex scalar components of $\F^i$, $i=1,2,\ldots,m$. 
Hence, eq.~(2.2) is just the N=1 supersymmetric extension of eq.~(2.3) whose NLSM target 
space is a K\"ahler manifold~\footnote{We ignore here possible global complications 
related to K\"ahler geometry.} of complex dimension $m$, with the function $K$ being the 
K\"ahler potential for the K\"ahler NLSM metric \cite{zum}
$$ g_{i\bar{j}}=\fracmm{\pa^2 K}{\pa\f^i\pa\bar{\f}^{\bar{j}}}~~.\eqno(2.4)$$
Eq.~(2.2) thus provides us with the manifestly supersymmetric and {\it universal\/} 
description of {\it all\/} 4d, N=1 supersymmetric NLSMs in terms of a single 
non-holomorphic real potential $K(\F,\Bar{\F})$. As is obvious from eq.~(2.2), 
the K\"ahler potential is defined modulo K\"ahler gauge transformations,
$$ K(\F,\Bar{\F})~\to~  K(\F,\Bar{\F})+f(\F) +\Bar{f}(\Bar{\F})~,\eqno(2.5)$$
with the holomorphic gauge parameter $f(\F)$. 

The {\it extended} supersymmetry in 4d NLSM is limited to N=2 since there exist no scalar 
supermultiplets beyond N=2 (by definition, all physical bosonic components of NLSM are 
scalars). The 4d, N=2 extended supersymmetry can be equivalently described as 6d, N=1 
supersymmetry which has the same number (8) of supercharges.~\footnote{See e.g.,  
ref.~\cite{sto} for a discussion of supersymmetric NLSM in 6d.}  An N=2 supersymmetric 
extension of a 4d NLSM (2.1) only exists if its metric is {\it hyper-K\"ahler}. This 
fact was initially established in components, by analyzing the restrictions imposed
by extended supersymmetry on a K\"ahler potential \cite{alfr}. The 4d, N=2 scalar 
multiplet is called {\it hypermultiplet} \cite{fs}. Though there exist many different
off-shell versions of a hypermultiplet in the conventional N=2 superspace (see e.g.,
refs.~\cite{nlsmbook,lkt} for some earlier references, or a recent paper \cite{kuzp}), 
the universal, manifestly N=2 supersymmetric formulation of the hypermultipet is only
possible in the N=2 {\it harmonic superspace} (HSS) introduced in 
ref.~\cite{gikos}.~\footnote{A brief introduction into HSS is given in Appendix A.}
The HSS is an extension of the conventional N=2 superspace by a two-sphere 
$S^2=SU(2)/U(1)$. Among the most important properties of HSS are (i) the existence of
invariant subspace called {\it analytic}, and (ii) the description of $SU(2)$ tensor
representations in terms of objects having definite $U(1)$ charge. Analytic N=2
superfields can be considered as the N=2 counterparts to N=1 chiral superfields, whereas
the harmonic calculus can be efficiently performed by employing isospinor harmonics 
$u^{\pm}_i$ instead of the usual polar coordinates $(\vf,\vq)$ on the sphere. 
Hypermultiplets in HSS are described by two basic types of unconstrained analytic 
superfields (usually denoted as $q$ and $\o$), which are dual to each other (see eqs.~(2.6) 
and (2.7) below) and can be chosen to be dimensionless. The physical components of a 
$q$-superfield comprise a complex scalar $SU(2)$-doublet and a Dirac spinor singlet. The 
physical components of an $\o$--superfield comprise a real scalar singlet, a scalar 
$SU(2)$-triplet and a chiral spinor $SU(2)$ doublet. A $q$-superfield is complex and has 
$U(1)$ charge one, whereas the real $\o$-superfield has vanishing $U(1)$ charge. The classical 
duality transformation between them reads \cite{gikos}
$$ q^{+}_a=u^{+}_a\o + u^-_af^{++}~,\eqno(2.6)$$
where yet another $SU(2)_{\rm PG}$ doublet $q^{+}_a=(q^+,\sbar{q}{}^{+})$, $a=1,2$, and the 
auxiliary analytic complex superfield $f^{++}$, which plays the role of a Lagrange multiplier, 
have been introduced. Inverting eq.~(2.6) yields
$$ \o=u^-_aq^{a+} \quad {\rm and}\quad f^{++}=-u^+_aq^{a+}~.\eqno(2.7)$$

The hyper-K\"ahler manifold is a 4n-dimensional Riemannian manifold $\cm$ whose holonomy 
group is a subgroup of $Sp(n)$. Directly imposing the hyper-K\"ahler condition on a 
K\"ahler potential results in the non-linear (Monge-Amper\'e) partial differential 
equation \cite{hkl}. The HSS offers a formal solution to this equation in the form of the
most general N=2 supersymmetric NLSM having the action 
$$ S[q,\sbar{q}]= \fracmm{1}{\k^2}\int d\z^{(-4)}du \,\cl^{(+4)}
(q^+,\sbar{q}{}^+,D^{++}q^+,D^{++}\sbar{q}{}^{+};u^{\pm}_i) \eqno(2.8)$$
over the analytic subspace of HSS whose measure $d\z^{(-4)}du$ is of $U(1)$ charge $(-4)$.
Here $D^{++}$ is the N=2 covariant harmonic derivative of dimension zero and of $U(1)$ 
charge $(+2)$. The analytic function $\cl^{(+4)}$ has to be of $U(1)$ charge $(+4)$ in 
order to compensate the opposite $U(1)$ charge of the measure, while it has to be  of
the first order in the derivatives $D^{++}q$ in order to guarantee the presence of the
standard NLSM kinetic term (2.1) in the corresponding component NLSM action, i.e. without 
higher spacetime derivatives. 

By N=2 supersymmetry eq.~(2.8) thus uniquely determines, in principle, the component 
hyper-K\"ahler NLSM metric in terms of a single analytic function $\cl^{(+4)}$. Their 
explicit relation is, however, highly non-trivial (and, in fact, not a 1-1 correspondence)
since eq.~(2.8) contains infinitely many auxiliary field components whose 
elimination requires solving inifinitely many linear differential equations on the sphere
altogether. This cumbersome procedure in HSS was only performed in a few special cases of N=2
NLSM with four-dimensional hyper-K\"ahler target spaces \cite{gios,giot,gi,kun}. 
Yet another caveat related to the infinite number of auxiliary fields is a considerable
redundancy of the HSS description of an N=2 NLSM, which exhibits itself in the existence 
of many apparently different analytic HSS lagrangians leading to the same hyper-K\"ahler 
metric in components (see sect.~3 for some explicit examples). To make the things more 
tractable, let's consider only those analytic lagrangians $\cl^{(+4)}$ that have a 
well-defined kinetic term, i.e. are of the form
$$ \cl^{(+4)}= -\sbar{q}{}^+D^{++}q^+ + \ck^{(+4)}(\sbar{q}{}^{+},q^+;u^{\pm})~,
\eqno(2.9)$$
where the analytic potential $\ck^{+(4)}$ is known as a {\it hyper-K\"ahler potential}
\cite{gios,giot}. 

Eq.~(2.9) naturally arises as the exact {\it low-energy effective action} (LEEA) for 
hypermultiplets in quantized N=2 supersymmetric gauge field theories \cite{ket1,ikz}.
An explicit dependence of the function $\ck$ upon harmonics signals the breaking of the
internal $SU(2)$ symmetry rotating two spinor charges of N=2 supersymmetry. Since the
duality relation (2.6) between $q$ and $\o$ hypermultiplets involves harmonics, it may
be useful to re-introduce a dependence upon both superfields into eq.~(2.9) if it results
in the absence of any explicit dependence upon harmonics. This is particularly relevant
in the context of the hypermultiplet LEEA  since the latter is normally dependent upon a 
dynamically generated real scale $\L$ which is interpreted as the expectaion value of 
some real Higgs hypermultiplet $\o$, 
$$\L=\VEV{\o}=const>0~.\eqno(2.10)$$

As is clear from eqs.~(2.8) and (2.9), a general hyper-K\"ahler metric does not have any 
isometries, and this is precisely the fact that makes its explicit construction via HSS
to be so difficult. It is to be compared with a derivation of hyper-K\"ahler metrics from
N=2 matter self-interaction in the conventional superspace \cite{lro,klr,nlsmbook}, which is
usually accompanied by duality transformations and leads to a presence of isometries in the 
hyper-K\"ahler metrics to be derived by using a finite number of auxiliary field components 
(see refs.~\cite{gid,kuzp} for a HSS reformulation of off-shell versions of hypermultiplet with 
a finite number of auxiliary fields). 
It is therefore the absence of isometries in a hyper-K\"ahler metric that is apparently 
responsible for the failure to formulate a manifestly N=2 supersymmetric NLSM with the same metric 
in the conventional N=2 superspace, i.e. with a finite number of N=2 auxiliary fields. Though HSS 
is capable of providing such an N=2 NLSM formulation in principle, the elimination of HSS 
auxiliary fields to recover a component metric of the N=2 NLSM without isometries 
represents a fundamental technical problem.

We are mostly going to restrict ourselves to four-dimensional (euclidean) hyper-K\"ahler 
NLSM target spaces having at least one isometry.  Since the Ricci tensor of a hyper-K\"ahler 
metric vanishes, any hyper-K\"ahler space is an Einstein space.~\footnote{A 
four-dimensional hyper-K\"ahler manifold can be equivalently characterized either as a
complex \newline ${~~~~~}$ K\"ahler and Ricci-flat (i.e Calabi-Yau) manifold, or as a 
real one with self-dual curvature.} A study of four-dimensional (euclidean) Einstein 
spaces having an isometry (i.e. with a metric to be independent upon one coordinate 
$(\rho)$ in some coordinates $\rho,y_i$, $i=1,2,3$) allows one to distinguish the 
hyper-K\"ahler spaces among the Einstein spaces by the following form of metric \cite{ghaw}: 
$$ ds^2_{[4]}= H(d\vec{y})^2+H^{-1}(d\rho +\vec{C}\cdot d\vec{y})^2~,\eqno(2.11)$$
where we have used the notation $\vec{y}=\{y_i\}$ and $\vec{\de}=\{\pa_i\}$. The vector 
function $\vec{C}(y_i)$ is supposed to satisfy the first-order equation
$$\vec{\de}\times\vec{C}=\vec{\de}H~,\eqno(2.12)$$
whereas the function $H(y_i)$ has to be harmonic, i.e. satisfy the Laplace equation
$$ \D H(\vec{y})=0\eqno(2.13)$$
outside the origin $\vec{y}=0$. It is worth mentioning that the dummy coordinate $\rho$ 
should be periodic (of period $2\p k$, $k\in {\bf Z}$) in order to avoid conical 
singularities in the metric (2.11).

An explicit relation between a harmonic function $H$ and a hyper-K\"ahler potential 
$\ck$ of the corresponding N=2 NLSM in HSS was established in ref.~\cite{gi}. One needs 
just a single $q$-hypermultiplet, having four real bosonic physical components, in order
to parameterize a four-dimensional hyper-K\"ahler NLSM target space, with an isometry 
being represented by a rigid $U(1)$ symmetry of the hyper-K\"ahler potential with respect to
the hypermultiplet rotations
$$ q^+\to e^{i\a}q^+~,\quad  \sbar{q}{}^+\to e^{-i\a}\sbar{q}{}^+~.\eqno(2.14)$$
This implies that the hyper-K\"ahler potential of $U(1)$ charge $(+4)$ is an analytic 
function of the invariant product $(q\!\sbar{q})$ of $U(1)$ charge $(+2)$, i.e.
$\ck=\ck(q\!\sbar{q},u)$. Hence, one has \cite{gi}
$$ \ck^{(+4)}=\sum^{\infty}_{l=0}\,\x^{(-2l)}\fracmm{(\sbar{q}{}^+q^+)^{l+2}}{l+2}~,
\eqno(2.15)$$
where the harmonic-dependent `coefficients' $\x^{(-2l)}(u)$ have been introduced,
$$ \x^{(-2l)}=\x^{(i_1\cdots i_{2l})}u^-_{i_1}\cdots u^-_{i_{2l}}~,\qquad l=1,2,\ldots~.
\eqno(2.16)$$
The latter are subject to the reality condition
$$ \sbar{\x}{}^{(-2l)}=(-1)^l\x^{(-2l)}~.\eqno(2.17)$$
 
A general solution to eq.~(2.13) reads 
$$ H=\fracmm{const.}{2r}+\fracmm{U(\vec{y})}{2}~,\eqno(2.18)$$
with the function $U(\vec{y})$ being non-singular in the origin.  Hence, the latter can be 
decomposed in terms of the standard momentum eigenfunctions 
$Y_l^m(\vq,\vf)$ depending upon the spherical coordinates $(r,\vq,\vf)$ with
$r=\sqrt{\vec{y}{}^{\,2}}$ as follows:
$$U(\vec{y})=\sum^{+\infty}_{l=0}\,\sum_{m=-l}^{m=+l}\,c_{lm}r^lY_l^m(\vq,\vf)~.\eqno(2.19)$$
The one-to-one correspondence between the integration constants $c_{lm}$ of eq.~(2.19) and the 
hyper-K\"ahler potential coefficients of eq.~(2.15) is given by~\cite{gi}
$$ \x^{i_1=1,\ldots,i_{l-m}=1,i_{l-m+1}=2,\ldots,i_{2l}=2} = \fracmm{c_{lm}}{C}
\fracmm{(2l+1)}{(l+1)}~,\eqno(2.20)$$
where $C$ is a normalization constant whose exact value is irrelevant for our purposes.

A physical meaning of the harmonic potential $H$  is transparent for the solitonic (regular)
{\it multicentre} hyper-K\"ahler metrics \cite{ghaw}, which are defined by 
$$H(\vec{y})=1+ \sum^p_{A=1} \fracmm{\abs{k_A}}{2\abs{\vec{y}-\vec{y}_A}}~~.\eqno(2.21)$$

The corresponding solution (2.11) to the euclidean Einstein equations describes $p\geq 1$ 
gravitational (Gibbons-Hawking) instantons, each having a topological charge $k_A\in{\bf Z}$ 
and `sitting' at a space point $\vec{y}_A$. Since all these solutions are actually independent 
upon the fourth coordinate $\rho$, they can also be interpreted as three-dimensional (static)
{\it multi-monopole} solutions with $4p$ moduli $\{k_A,\vec{y}_A\}$. They are also known as the
{\it Kaluza-Klein} (KK) monopoles in the literature \cite{kkm1,kkm2}. Though the HSS moduli 
$\x^{(i_1\cdots i_{2l})}$ in the alternative HSS description of the same multi-monopole 
configuration have no direct physical interpretation and they are not independent at all, the 
HSS description itself in terms of the analytic hyper-K\"ahler potential (2.15) is manifestly 
{\it non-singular}. The latter is useful in M-theory and brane technology, when describing the 
gauge symmetry enhancement for coincident D-6-branes in non-singular terms (sect.~5). Note that
the BPS nature of a KK monopole means that its mass is equal to its charge (in dimensionless
units). 

A non-vanishing central charge of N=2 supersymmetry algebra can be easily incorporated into
the HSS formalism by modifying the harmonic covariant derivative $D^{++}$. It simply amounts to 
introducing into NLSM a minimal coupling of hypermultiplets with an N=2 abelian background gauge
superfield having the constant N=2 superfield strength equal to the central charge (Appendix A).
As was shown in refs.~\cite{kun,ikz}, a non-vanishing central charge
leads to the appearance of a non-trivial scalar potential in components, whose form is entirely
determined by a hyper-K\"ahler metric of the kinetic NLSM terms. This fact will play an important
role in the mechanism of dynamical generation of N=2 vector multiplets in N=2 NLSM (sect.~4). 

\section{N=2 NLSM with ALE metric}

Let's take the harmonic potential (2.21) describing the two-centered $(p=2)$ monopole solution 
with equal charges $(k_A=1,~\vec{y}_1=\vec{0},~\vec{y}_2=\vec{\x}$), and modify it by a constant
$\l>0$ as
$$ H(\vec{y})=\l + \fracmm{1}{2} \left\{ \fracmm{1}{\abs{\vec{y}-\vec{0}}} +
\fracmm{1}{\abs{\vec{y}-\vec{\x}}}\right\}~.\eqno(3.1)$$ 
The real vector $\vec{\x}$ can be equally represented as an $SU(2)$ triplet 
$\x^{ij}=i\vec{\x}\cdot\vec{\t}^{ij}$ satisfying the reality condition
$$ (\x^{ij})^{\dg}\equiv\x^{\dg}_{ij}=\ve_{il}\ve_{jm}\x^{lm}=\x_{ij}~,\eqno(3.2)$$
where $\vec{\t}$ are the usual $2\times 2$ Pauli matrices. The hyper-K\"ahler metric defined by 
eqs.~(2.11) and (3.1) is called the {\it double Taub-NUT} metric with a constant potential $\l$ 
at spacial infinity \cite{ath}. In accordance with the general results of sect.~2, the N=2 NLSM 
with the same target space metric is described by the HSS Lagrangian \cite{gi}
$$ \cl^{(+4)}=-\sbar{q}{}^{+}_A D^{++}q^+_{A} - V^{++}\left( \ve^{AB}\sbar{q}{}^{+}_Aq^+_{B} 
+\x^{++}\right)-\l\left(\sum_{A=1}^2 \sbar{q}{}^+_Aq^+_A\right)^2~,\eqno(3.3)$$
where the N=2 vector gauge superfield $V^{++}$ has been introduced as a Lagrange multiplier,
and $\x^{++}=\x^{ij}u^+_iu^+_j$. As is clear from eq.~(3.3), this NLSM is invariant under
the local $U(1)$ gauge symmetry
$$\d q^+_1=\L q^+_2,~\quad \d q^+_2=-\L q^+_1,~\quad \d V^{++}=D^{++}\L~,\eqno(3.4)$$
with the analytic HSS superfield parameter $\L(\z,u)$. The rigid $SU(2)$ automorphisms of N=2 
supersymmetry algebra are obviously broken in eq.~(3.3) to its abelian subgroup that leaves
$\x^{++}$ invariant. The extra (Pauli-G\"ursey) symmetry $SU(2)_{\rm PG}$ rotating $q$ and
$\sbar{q}$~ is also broken in eq.~(3.3) unless $\l\neq 0$. 

Eq.~(3.3) takes a particularly simple form in the limit $\x\to 0$ where it reduces (after a
superfield redefinition) to the well-known {\it Taub-NUT} NLSM action in HSS \cite{gios}. 
Similarly, in another limit $\l\to 0$, eq.~(3.3) yields the N=2 NLSM with the {\it Eguchi-Hanson}
(EH) metric \cite{giot}. In other words, the double Taub-NUT metric interpolates between the
Taub-NUT and Eguchi-Hanson metrics \cite{gi}, as was also explicitly demonstrated in 
ref.~\cite{gi}. In both limits (Taub-NUT and Eguchi-Hanson), the metric has $U(2)$ isometry, 
whereas only a $U(1)$ isometry is left when both $\l\neq 0$ and $\x\neq 0$.

Eq.~(3.3) at $\l=0$ takes the form of the $SU(2)_{\rm PG}$-invariant minimal coupling between the
two `matter' FS-type hypermultiplets $q^+_A$ and an abelian N=2 vector gauge multiplet $V^{++}$
in the presence of a gauge-invariant (electric) {\it Fayet-Iliopoulos} term linear in $V^{++}$,
$$ \cl^{(+4)}(q_A,V)=-\frac{1}{2}q^{a+}_A D^{++}q^+_{aA} - V^{++}\left(\frac{1}{2}
\ve^{AB}q^{a+}_Aq^{+}_{Ba}+\x^{++}\right)~.\eqno(3.5)$$
This HSS Lagrangian can be rewritten after some algebra to the following (classically equivalent)
form \cite{giot}:
$$ \cl^{(+4)}(q)
=-\frac{1}{2}q^{a+}D^{++}q^+_a +\fracmm{(\x^{++})^2}{2(q^{a+}u^-_a)^2}~~,\eqno(3.6)$$
which determines the hyper-K\"ahler potential of the EH metric according to eq.~(2.9). After
the duality transformation (2.6), one arrives at the dual action \cite{giot}
$$ \cl^{(+4)}(\o)= -\frac{1}{2}(D^{++}\o)^2 +\fracmm{(\x^{++})^2}{2\o^2}\eqno(3.7)$$
in terms of the single real $\o$ superfield. Therefore, the N=2 supersymmetric NLSM Lagrangian 
in HSS for a given (in this case, Eguchi-Hanson) hyper-K\"ahler metric is not unique. 

Let's now define yet another gauge-invariant HSS action in terms of another two FS-type 
hypermultiplet superfields and an N=2 vector gauge $V^{++}$ superfield as 
$$ S_{\rm EH}[q_1,q_2,V]=\int d\z^{(-4)}du \left[ -\sbar{q}{}^+_1\cd^{++}q_1^+
+ \sbar{q}{}^+_2\cd^{++}q_2^+ + V^{++}\x^{++}\right]~,\eqno(3.8)$$
where we have returned to canonical dimensions for all the superfields involved,~\footnote{
In units of mass one has $[q]=1$, $[V]=0$ and $[\x]=+2$.} and introduced the gauge-covariant 
harmonic derivative \cite{gikos}
$$ \cd^{++}=D^{++}+iV^{++}~,\eqno(3.9)$$
thus extending the rigid $U(1)$ symmetry (2.14) of a free hypermultiplet action to the local 
analytic one. It is not difficult to check that the classical theory (3.8) is equivalent to 
that of eq.~(3.5), e.g. by considering a gauge $q^+_2=0$ in eq.~(3.8) and a gauge $q^+_2=iq^+_1$
in eq.~(3.5), up to rescaling by a factor of 2. However, in the form (3.8), the 
$SU(2)_{\rm PG}$ invariance is no longer manifest. Moreover, the action (3.8) has the wrong 
sign in front of the kinetic term for the $q_2$ hypermultiplet that indicates its non-physical 
(ghost) nature. This also implies its anti-causal propagation and the wrong (negative) sign of 
the residue in the propagator of $q_2$ superfield (see the next sect.~4). It does not, however, 
make our theory (3.8) non-unitary since the $q^+_2$ hypermultiplet is a gauge degree of freedom,
while the classical action (3.8) itself is dual to any of the manifestly unitary NLSM actions 
with the ALE (Eguchi-Hanson) target space in eqs.~(3.6) and (3.7). The action (3.8) has the form 
of a non-compact gauged NLSM over the coset $SU(1,1)/U(1)$ parameterized by the FS-type 
hypermultiplets $q_A$ in the fundamental representation of $SU(1,1)$ whose $U(1)$ subgroup is 
gauged in HSS. 

We are going to exploit the freedom of choosing a classical HSS Lagrangian with the on-shell
Eguchi-Hanson metric and to take eq.~(3.8) as our starting point for quantizaton. It is worth 
mentioning here that the minimal gauge interaction of hypermultiplets with N=2 vector multiplets 
is the {\it only\/} renormalizable type of N=2 supersymmetric field-theoretical interaction in 
four-dimensional spacetime \cite{hst}.
The classical correspondence with the formally unitary (but non-renormalizable) NLSM actions 
(3.6) and (3.7) ensures unitarity in our theory, whereas the non-anomalous gauge Ward identities
should take care of the gauge invariance after quantization. Our approach may be compared to the 
standard bosonic string theory where the 2d Nambu-Goto classical string action \cite{ng} is 
substituted by the 2d Polyakov string action \cite{pols}. The non-polynomial Nambu-Goto action 
has a clear geometrical interpretation as the area of a string world-sheet but it is formally 
non-renormalizable. One defines a quantized bosonic string theory (in the critical dimension) 
after replacing the Nambu-Goto action by the classically equivalent Polyakov action which has 
the 2d auxiliary metric as a Lagrange multiplier. In our case, however, we will not integrate 
over our Lagrange multiplier given by an N=2 vector gauge superfield in 4d. The ghost 
hypermultiplet will be integrated out in quantum theory (sect.~4).

The quantized theory (3.8) is, however, of little interest unless it is supplemented by an
N=2 central charge $\hat{Z}$ giving BPS masses to hypermultiplets. A hypermultiplet of mass $m$
can be described in HSS via the extension \cite{gikos} 
$$ \hat{D}^{++}\equiv D^{++} +i(\q^{\a +}\q^{+}_{\a})\hat{\bar{Z}} +
i(\bar{\q}^{+}_{\dt{\b}}\bar{\q}^{\dt{\b}+})\hat{Z} \eqno(3.10)$$
of the flat harmonic derivative $D^{++}$, with $\hat{Z}$ being an operator. It is is not
difficult to verify (see, e.g. ref.~\cite{ikz}) that the free hypermultiplet equation 
of motion $\hat{D}^{++}q^+=0$ implies 
$$ \left( \bo +\hat{Z}\hat{\bar{Z}} \right) q^+=0~,\eqno(3.11)$$
which allows us to identify $\hat{Z}\hat{\bar{Z}}q^+=m^2q^+$. Because of eq.~(3.9), the
modification (3.10) in the case of a single charged hypermultiplet amounts to adding a minimal 
coupling to the particular N=2 abelian vector gauge superfield background having the constant 
N=2 gauge superfield strength equal to the central charge value \cite{bbiko,ikz} (see also 
Appendix A). We are going to use the original interpretation (3.10) of the central charge, by
associating it to $D^{++}$ not $V^{++}$, since we can then introduce {\it different} masses 
for the hypermultiplets $q_1^+$ and $q_2^+$ in eq.~(3.8) via eqs.~(3.9) and (3.10). 

N=2 central charges in 4d HSS can be generated from a 6d HSS by the use of the standard
(Scherk-Schwarz) mechanism of dimensional reduction \cite{ssw}, where the derivatives with 
respect to extra space coordinates play the role of the central charge operators (see e.g.,  
refs.~\cite{giot,ikz} for details).  As was noticed in ref.~\cite{giot}, the six-dimensional
notation may sometimes simplify the equations with implicit central charges. For example, the
bosonic kinetic terms of the NLSM (3.8) to be rewritten to 6d, after elimination of the HSS 
auxiliary fields in components, are given by \cite{giot,kun}
$$ \eqalign{
S_{\rm bosonic}[\f^{ai}_1,\f^{ai}_2,V_{\m}]~=~& 
\frac{1}{2}\int d^6x \left\{ (D^{\m} \f^{ia}_1) (D_{\m}\f_{ia1})
- (D^{\m} \f^{ia}_2) (D_{\m}\f_{ia2}) \right. \cr
~&~~~~~~~~~~~~~~~ \left. +\frac{1}{2}D_{ij} 
\left( \f^{ia}_1\f^j_{1a}-\f^{ia}_2 \f^j_{2a}+\x^{ij} \right) \right\},\cr}\eqno(3.12)$$
where $\m=0,1,2,3,4,5$, $D_{\m}=\pa_{\m}+iV_{\m}$, and $D_{ij}$ is the scalar triplet of the 
auxiliary field components of the N=2 vector superfield $V^{++}$ in a WZ-gauge.

We are now in a position to formulate our model by the following HSS action:
$$\eqalign{
S_{\rm ALE} [q_1,q_2,V] =  & \int d\z^{(-4)}du \left\{ 
-\sbar{q}{}^+_1(\hat{D}^{++}_1+iV^{++})q_1^+ \right. \cr
&~~ \left. + \sbar{q}{}^+_2(\hat{D}^{++}_2+iV^{++})q_2^+ + V^{++}\x^{++}\right\}~,\cr}
\eqno(3.13)$$
where 
$$\hat{Z}\hat{\bar{Z}}q^+_1=m^2_1q^+_1~,\quad \hat{Z}\hat{\bar{Z}}q^+_2=m^2_2q^+_2~.\eqno(3.14)
$$
It should be remembered that the mass parameters $m^2_1$ and $m^2_2$ introduced in eq.~(3.14) do
not represent physical masses. As is clear from eq.~(3.12), the classical on-shell physical 
significance has only their difference
$$ m^2_2-m^2_1\equiv m^2~,\eqno(3.15)$$
which can be identified with the classical mass of the single physical hypermultiplet in the 
NLSM under consideration, after taking into account the constraint imposed by the Lagrange 
multiplier $D_{ij}$ ({\it cf.} ref.~\cite{vanh}). Moreover, because of the presence of a FI
term linear in $D_{ij}$ in the action (3.12), the auxiliary triplet $D^{ij}$ of $V^{++}$ may 
develop a non-trivial vacuum expectation value in quantum theory after taking into account
quantum corrections due to quantized hypermultiplets This will influence the physical mass 
values to be defined with respect to a `true' vacuum. 

Accordingly, we first have to examine in the next sect.~4 whether the auxiliary field components
$D_{ij}$ get a non-trivial vacuum expectation value. The latter has to be constant in order to 
maintain 4d Lorentz invariance. A constant solution $\VEV{D_{ij}}\neq 0$ is clearly consistent
with the abelian gauge invariance in components (Appendix B).

\section{Quantum theory}

To quantize both hypermultiplets of the theory (3.8) in a manifestly N=2 supersymmetric way, 
we need a quantum perturbation theory in terms of analytic HSS superfields in four spacetime 
dimensions. The HSS Feynman rules for massless N=2 supersymmetric gauge field theories were 
first obtained in ref.~\cite{gikos}. For our purposes in subsect.~4.2, we use a massive 
hypermultiplet propagator which was first derived in ref.~\cite{zupnik}. In subsect.~4.3 we 
employ its generalization depending upon a background FI term too \cite{ikz}. A HSS propagator of 
the unphysical hypermultiplet has some important differences in comparison to the physical 
hypermultiplet propagator, which are discussed in subsect.~4.1 along the lines of ref.~\cite{vanh}. 
A manifestly N=2 supersymmetric derivation of the low-energy gauge effective action (LEEA) to be 
obtained by an integration over a single matter hypermultiplet minimally coupled to an N=2 
abelian vector gauge superfield in HSS is discussed at length in ref.~\cite{bbiko} (see also 
refs.~\cite{div,seib} for some earlier component results, and some recent papers 
\cite{myrev,ketpr} about a relation between HSS and components). These results are used in 
subsect.~4.3 to argue for a dynamical generation of an N=2 vector multiplet in the 4d quantum field 
theory (3.8). Further evidence coming from M-theory and brane technology is discussed in the next 
sect.~5, whereas some generalizations are outlined in sect.~6

\subsection{HSS propagators for hypermultiplets}

The physical HSS propagator for a massive FS-type hypermultiplet reads \cite{zupnik,ikz}
$$ i\VEV{q^+(1)\sbar{q}{}^+(2)}_{\rm phys} = 
\fracmm{-1}{\bo + m^2-i0}(D^+_1)^4(D^+_2)^4e^{v_2-v_1}\d^{12}(Z_1-Z_2)
\fracmm{1}{(u_1^+u^+_2)^3}~,\eqno(4.1)$$
where $v$ is the so-called `{\it bridge}' defined by the general rule 
$$ \cd= e^{-v}D e^{v} \eqno(4.2)$$
between the manifestly analytic HSS derivatives $\cd$ and the covariantly analytic ones 
$D$. In the case of the central charge background (3.10) one easily finds \cite{ikz}  
$$ v=i(\q^+\q^-)\hat{\bar{Z}}+ i(\bar{\q}^+\bar{\q}^-)\hat{Z}~.\eqno(4.3)$$
The Green function $G^{(1,1)}(1|2)_{\rm phys}\equiv i\VEV{q^+(1)\sbar{q}{}^+(2)}_{\rm phys}$ 
satisfies the equation
$$ \hat{D}^{++}_1G_{\rm phys}^{(1,1)}(1|2)=\d^{(3,1)}_{\rm A}(1|2)~,\eqno(4.4)$$
where the analytic HSS delta-function $\d^{(3,1)}_{\rm A}(1|2)$ has been introduced \cite{gikos}.

A causal (unitary) propagation is ensured in quantum field theory by adding a small negative 
imaginary part to the mass squared, $m^2\to m^2-i\e$, in the propagator (4.1) \cite{wein}. The
same prescription automatically takes care of (i) the convergence of the path integral defining 
the generating functional of quantum Green's functions in Minkowski spacetime and (ii) free 
interchange of integrations. A propagator of the non-physical hypermultiplet entering the action
(3.8) with the wrong sign (and, hence, formally leading to negative norms of the corresponding 
`states') is also of the form (4.1) but with the negative residue 
{\it and\/} an anti-causal $i\e$-prescription (Fig.~1), 
$$ i\VEV{q^+(1)\sbar{q}{}^+(2)}_{\rm nonphys} = 
\fracmm{1}{\bo + m^2+i0}(D^+_1)^4(D^+_2)^4e^{v_2-v_1}\d^{12}(Z_1-Z_2)
\fracmm{1}{(u_1^+u^+_2)^3}~.\eqno(4.5)$$
It can only occur as an internal line inside Feynman graphs, similarly to the HSS ghost
hypermultiplet propagators in N=2 supersymmetric gauge field theories considered in 
ref.~\cite{bbiko}. Our quantized non-physical hypermultiplet has, however, bosonic statistics.

\begin{figure}
\vglue.1in
\makebox{
\epsfxsize=4in
\epsfbox{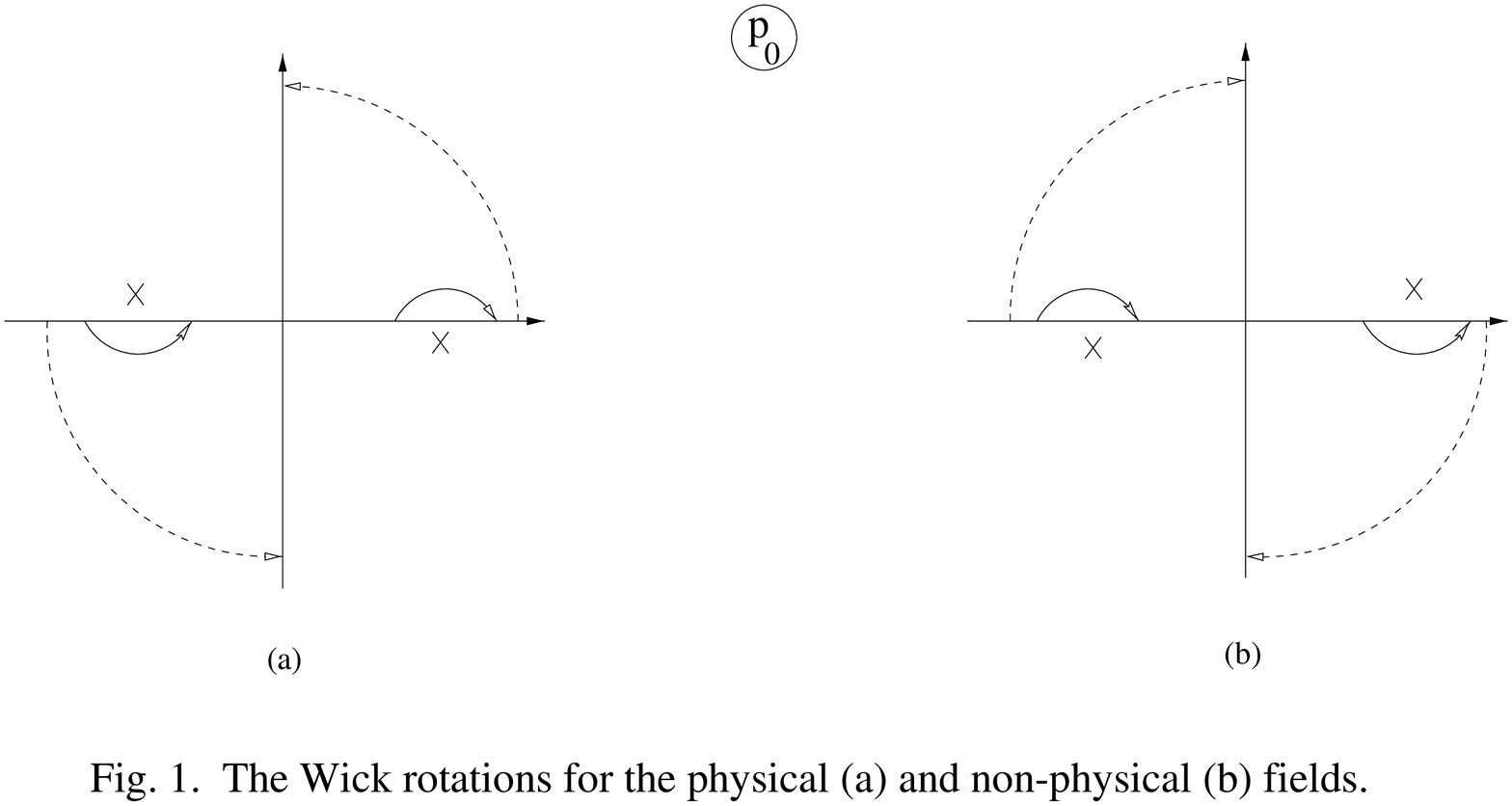}
}
\end{figure}

Gauge couplings of physical hypermultiplets to N=2 vector superfields also differ by minus sign 
from those of non-physical hypermultiplets. Hence, a Feynman graph for the non-physical fields
has extra minus signs for every internal line and every vertex, when being compared to the same 
graph for the physical fields. Though all these signs mutually cancel in loop diagrams with the 
same number of vertices and internal lines, the difference in $i\e$ prescription remains. It
forces the non-physical poles in the complex $p_0$-plane to be on the other side of the real 
axis \cite{vanh}. This amounts to the appearance of a relative minus sign for every non-physical
loop compared to the same physical loop, because of the opposite Wick rotation in the momentum 
space for the non-physical fields (Fig.~1). In this respect, the quantized non-physical 
(or of negative-norm) fields behave like fermions or Pauli-Villars regulators, so that one 
may already expect UV-divergence cancellations in Feynman graphs between physical and 
non-physical loops. It happens to be the case indeed (see the next subsections).

\subsection{Gauge LEEA and vacuum structure}

We are now in a position to discuss the N=2 gauge low-energy effective action (LEEA) to be 
defined by a Gaussian integration over both hypermultiplets in eq.~(3.8) and then expanding the 
result in powers of external momenta. The quantum effective action $\G(V^{++})$ is formally 
defined in HSS by the one-loop formula
$$ \G(V^{++})=i\Tr\ln \cd^{++}_{\rm phys}- i\Tr\ln \cd^{++}_{\rm nonphys}~,\eqno(4.6a)$$
or, in terms of the Green functions (4.4), as 
$$ \G(V^{++})= i\Tr\ln
\fracmm{\d_{\rm A}^{(3,1)}+iV^{++}G^{(1,1)}_{\rm phys}}{\d_{\rm A}^{(3,1)}
+iV^{++}G^{(1,1)}_{\rm nonphys}}~~~.\eqno(4.6b)$$
A supergraph calculation of the LEEA for a single physical hypermultiplet in HSS (Fig.~2) was
already done in refs.~\cite{bbiko,bt}, so that we can use the known results here. 

\begin{figure}
\vglue.1in
\makebox{
\epsfxsize=4in
\epsfbox{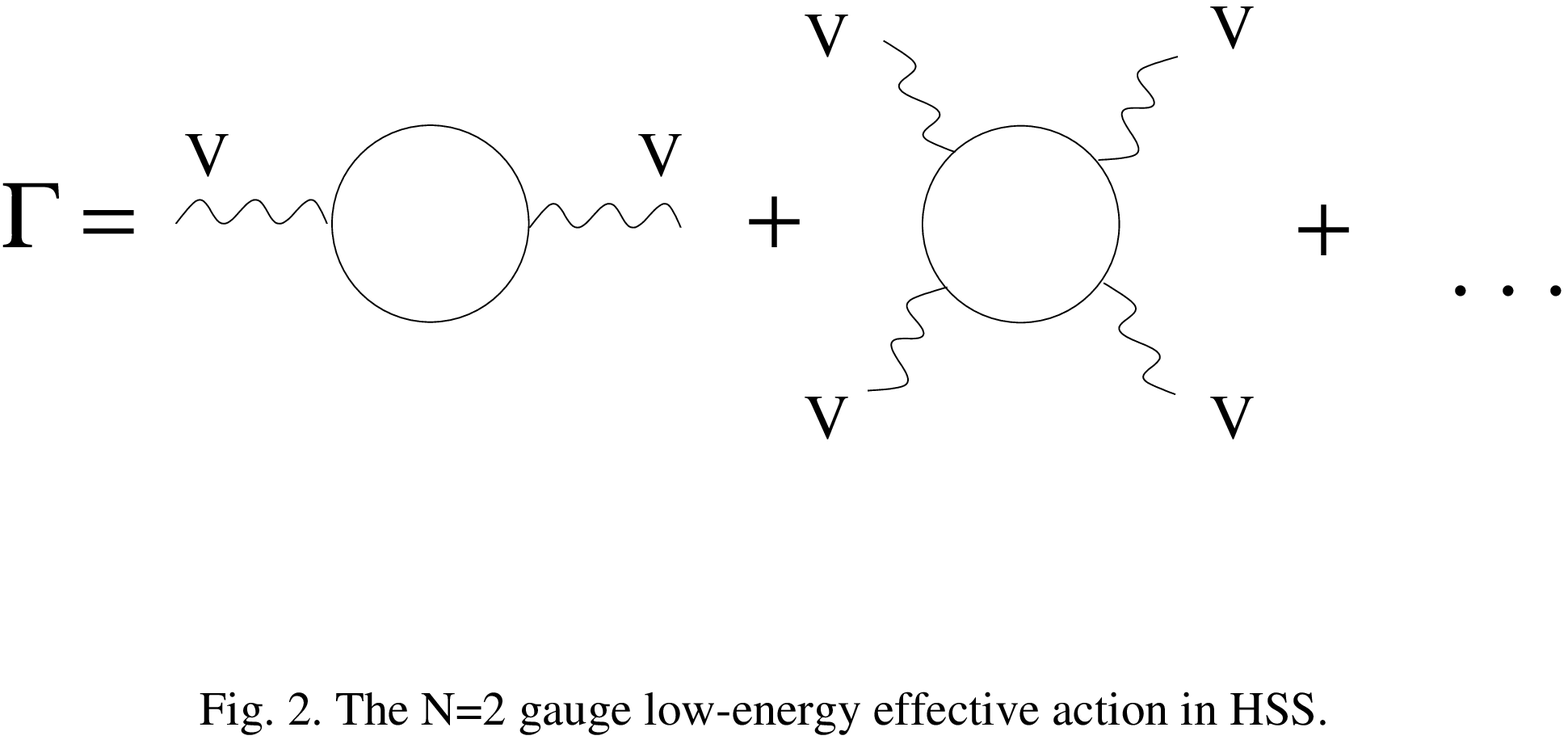}
}
\end{figure}

Because of the gauge invariance, $\G(V^{++})$ can only depend upon the abelian N=2 gauge 
superfield strength $W$ and its conjugate $\bar{W}$, with both being defined by eq.~(A.25). 
On dimensional reasons, the general structure of the gauge LEEA (modulo terms explicitly
depending upon N=2 superspace derivatives of $W$ or $\bar{W}$) is given by
$$ \G[V^{++}]=\left[ \int d^4xd^4\q \,\cf(W) +{\rm h.c.}\right] +\int d^4xd^4\q
d^4\bar{\q}\,\ch(W,\bar{W})~,\eqno(4.7)$$
where the leading {\it holomorphic} term is known as the (perturbative) Seiberg-Witten LEEA 
\cite{sw} or the integrated N=2 supersymmetric (chiral) $U(1)_R$ anomaly \cite{div,seib,myrev},
whereas the second term is called the {\it non-holomorphic} (perturbative) next-to-leading-order
correction \cite{ketpr}. It is worth mentioning that the real function $\ch(W,\bar{W})$ is
subject to the gauge transformations
$$ \ch(W,\bar{W})\to  \ch(W,\bar{W}) +F(W) +\bar{F}(\bar{W})~,\eqno(4.8)$$
with the N=2 {\it chiral} superfield parameter $F(W)$, quite similarly to the K\"ahler 
transformations (2.5) in N=1 superspace. The HSS calculations \cite{bbiko,ketpr} yield
$$ \cf(W)_{\rm phys}= -\fracmm{1}{(8\p)^2}W^2\ln\fracmm{W^2}{m^2}~,\eqno(4.9)$$
and
$$ \ch(W,\bar{W})_{\rm phys}=\fracmm{1}{(16\p)^2} \left(\ln\fracmm{W}{\L}\right)
\left(\ln\fracmm{\bar{W}}{\L}\right)~,\eqno(4.10)$$
where $\L$ is an irrelevant parameter since the action (4.7) does not really depend on it
because of eqs.~(4.8) and (4.10) \cite{ds}.

Eqs.~(4.9) and (4.10) in our case (4.6) immediately imply that
$$ \G[V^{++}]_{\rm LEEA} = 
-\fracmm{1}{32\p^2}\ln\left(\fracmm{m_2^2}{m^2_1}\right)\int d^4x d^4\q\,W^2
\equiv -\fracmm{1}{2e^2_0}\int d^4x d^4\q\,W^2~,\eqno(4.11)$$
which is just a free action of the N=2 vector gauge superfield~! Eq.~(4.11) implies, in 
particular, the dynamical generation of the term quadratic in the auxiliary field $D_{ij}$, 
which is accompanying the standard kinetic terms of the N=2 vector multiplet in the well-known 
component form (B.8) of eq.~(4.11). Together with the FI-term in eq.~(3.8), this now implies a 
non-vanishing vacuum expectation value $\VEV{D_{ij}}\neq 0$.

Some comments are in order.

Unlike the N=2 gauge LEEA for a single hypermultiplet (Fig.~2), our LEEA (4.6) is both infra-red
(IR) {\it and\/} ultra-violet (UV) finite. IR divergences are obviously absent due to 
non-vanishing hypermultiplet masses acting as IR-regulators. As regards the UV divergences of
the N=2 gauge LEEA for a physical hypermultiplet, the leading 2-point contribution in Fig.~2 is
known to be the only divergent one (all the higher $n$-point contributions in Fig.~2 are 
automatically UV finite on dimensional reasons) \cite{bbiko}. The holomorphic 2-point 
contribution to the N=2 gauge LEEA due to a single physical hypermultiplet reads \cite{bbiko}
$$\G^{(2)}_{\rm phys}[V]=-\fracmm{i}{2}\fracmm{1}{(2\p)^4}\int d^4p d^8\q du\,
V^{++}(p,\q,u,)\P_{\rm phys}(-p^2)V^{--}(-p,\q,u)~,\eqno(4.12)$$
where the (dimensionally regularized) one-loop structure function $\P_{\rm phys}(-p^2)$ has been 
introduced (with $\m$ as a renormalization scale), 
$$\P_{\rm phys}(-p^2)
=\m^{2\ve}\int_{\rm E} \fracmm{d^{4-2\ve}l}{(2\p)^{4-2\ve}} \fracmm{1}{[l^2+m^2]
[(l-p)^2+m^2]}~~.\eqno(4.13)$$
Eq.~(4.13) is logarithmically UV-divergent in four spacetime dimensions $(\ve\to +0)$. This UV
divergence is simultaneously the origin of the renormalization scale dependence of the 
renormalized low-energy effective action (4.7) via its holomorphic (anomalous) conribution. In 
our case (4.6), the UV divergence of the self-energy integral (4.13) cancels against the opposite 
UV divergence of the similar contribution to the LEEA due to the nonphysical hypermultiplet, 
{\it viz.}
$$\eqalign{
\P(-p^2) ~\equiv ~& \P_{\rm phys}(-p^2) + \P_{\rm nonphys}(-p^2) \cr
~& =\int_{\rm E} \fracmm{d^4l}{(2\p)^4}\left\{ \fracmm{1}{[l^2+m^2_1] [(l-p)^2+m^2_1]}
-\fracmm{1}{[l^2+m^2_2] [(l-p)^2+m^2_2]}\right\} \cr
~&= \fracmm{1}{16\p^2}\int^1_0 dx \ln\fracmm{m^2_2+p^2x(1-x)}{m^2_1+p^2x(1-x)}~~,\cr}
\eqno(4.14)$$
where Feynman parameterization has been used to evaluate the momentum integral in the euclidean 
domain (we assume that $m^2_2>m_1^2$). A continuation to Minkowski space entails changing the 
sign of $p^2$ --- this explains our notation, $\P(s)$ and $s=-p^2$, above. The function $\P(s)$
is analytic in the cut $s$ plane whose analytic structure is best exhibited by dispersion 
relations \cite{brown}. 

One obtains eq.~(4.11) from eqs.~(4.12) and (4.14) in the low-energy limit $p^2\to 0$. We took
the vanishing momenta since we are first interested in finding a Poincar\'e- and gauge-invariant
vacuum background solution. It can only be represented by a spacetime-independent N=2 vector
gauge superfield strength $\VEV{W}$ having the form
$$ \VEV{W}=\VEV{a} +\frac{1}{2}(\q^{\a}_i\q_{\a j})\VEV{D^{ij}}~,\eqno(4.15)$$
where merely constant vacuum expectation values of the bosonic scalar components of $W$ have been 
kept. We can assume that $\VEV{a}=0$ without a loss of generality since: (i) there is no equation
on $\VEV{a}$ at all, and (ii) a constant $\VEV{a}$ would simply amount to the {\it equal} shift 
of both hypermultiplet masses in the theory (3.8). Hence, we are left with the induced scalar 
potential 
$$ V(\vec{D})=\vec{\x}\cdot \vec{D}-\fracmm{1}{2e^2_0}\vec{D}^2~,\eqno(4.16)$$
in components, which has the only vacuum solution 
$$ \VEV{\vec{D}}=e^2_0\vec{\x}\neq 0~.\eqno(4.17)$$

It also follows from eqs.~(3.15), (4.11) and (4.17) that
$$ m^2_1=\fracmm{m^2}{e^{16\p^2/e^2_0}-1}~,\quad 
   m^2_2=\fracmm{m^2}{1-e^{-16\p^2/e^2_0}}~~.\eqno(4.18)$$

This simple exercise can also be repeated in HSS, by using the results of Appendix A. Varying 
the N=2 gauge effective action with respect to the abelian N=2 vector gauge superfield $V^{++}$ 
in HSS yields the equation of motion (in vacuum) 
$$ \fracmm{1}{e^2_0}(D^+)^4\VEV{A^{--}}=\x^{++}~,\eqno(4.19)$$
where the HSS potentials $A^{--}$ and $V^{++}$ are related via eq.~(A.21). A Poincar\'e-invariant
solution to eq.~(4.19) reads
$$ \VEV{V^{++}}=(\theta^+)^2(\bar{\theta}^+)^2 e^2_0\x^{--}~,\eqno(4.20)$$
and it is equivalent to eq.~(4.17) because of eq.~(A.29).

Any other non-trivial N=2 gauge LEEA (4.7) having the form different from that of eq.~(4.11), i.e. 
with a non-quadratic holomorphic function $\cf(W)$, does not admit a constant non-vanishing 
solution for $\vec{D}$, because of the appearance of an extra equation 
$\left.\pa^3\cf/\pa W^3\right|_{W=a}\vec{D}^2=0$ \cite{iz}. A non-vanishing 
value of $\VEV{\vec{D}}$ implies the appearance of Goldstone fermions (see the second line of 
eq.~(B.7) in Appendix B) which inhomogeneously transform under on-shell N=2 supersymmetry. 
In other words, the N=2 supersymmetry is {\it spontaneously broken\/} in our theory (3.8).

\subsection{Dynamical generation of composite particles}

As was shown in the preceeding subsect.~4.2, quantum effects due to hypermultiplets lead to
the appearance of the {\it propagating} (physical) abelian N=2 vector multiplet $V^{++}$. In the
classical theory (3.8), $V^{++}$ is merely present as a (non-propagating) Lagrange multiplier.
Because of eqs.~(4.12) and (4.14), the induced gauge coupling constant is momentum-dependent,
$$ \fracmm{1}{e^2_{\rm ind}}=\fracmm{1}{16\p^2}\int^1_0dx\,\ln\fracmm{m^2_2+p^2x(1-x)}{m^2_1
+p^2x(1-x)}=\fracmm{1}{e^2_0}+O(p^2/m^2)~.\eqno(4.21)$$
Notably, the UV finiteness enjoyed by our theory in four spacetime dimensions is also necessary 
for its consistency: if there were UV divergent contributions to $e^2_{\rm ind}$, they would 
have to be removed by the corresponding counterterm proportional to the N=2 gauge action. The
latter must, however, be absent in the bare action (3.8) since, otherwise, it would contradict
the classical nature of $V^{++}$ as a Lagrange multiplier.

In order to calculate the full gauge LEEA, one has to repeat a calculation of the HSS graphs
depictured in Fig.~2 in terms of the {\it new\/} hypermultiplet propagators to be defined with
respect to the `true' vacuum with the non-vanishing FI term (4.17). 

The Green function of a physical hypermultiplet in a generic N=2 vector superfield background 
$\hat{V}^{++}$ satisfies the defining equation
$$ \cd^{++}_1 G_{{\rm phys},\hat{V}}^{(1,1)}(1|2)=\d^{(3,1)}_{\rm A}(1|2)~,\eqno(4.22)$$
whose solution can be formally written down in the form 
$$  G_{{\rm phys},\hat{V}}^{(1,1)}(1|2) =
\fracmm{-1}{\bo_{\rm cov}-i0}(D^+_1)^4(D^+_2)^4e^{\cv_2-\cv_1}\d^{12}(Z_1-Z_2)
\fracmm{1}{(u_1^+u^+_2)^3}~~,\eqno(4.23)$$
where the covariantly constant `bridge' $e^{-\cv}$ and the covariant d'Alambertian 
$\bo_{\rm cov}$ in the analytic HSS have been introduced \cite{bbiko,ikz}. The defining
equation for the `bridge' reads
$$ \cd^{++}e^{-\cv}=(D^{++}+i\hat{V}^{++})e^{-\cv}=0~,\eqno(4.24)$$
whereas the defining equation for the covariant d'Alambertian is given by
$$ -\ha (D^+)^4(\cd^{--})^2 \F^{(p)}=\bo_{\rm cov}\F^{(p)}~,\eqno(4.25)$$
where $\F^{(p)}$ is a HSS analytic superfield of (positive) $U(1)$ charge $p$. The definition
(4.25) obviously implies that 
$$ \[ D^+_{\a},\bo_{\rm cov}\]= \[ \bar{D}^+_{\dt{\a}},\bo_{\rm cov}\]=0~.\eqno(4.26)$$
An explicit form of the operator $\bo_{\rm cov}$ in a generic background $\hat{V}^{++}$ 
was calculated in ref.~\cite{bbiko} in the covariantly analytic form
$$\eqalign{
\bo_{\rm cov.~analytic} = & \cd^{\m}\cd_{\m}+\frac{i}{2}(\cd^{\a+}W)\cd^{-}_{\a}
+\frac{i}{2}(\bar{\cd}^{+}_{\dt{\a}}\bar{W})\bar{\cd}^{\dt{\a}-} \cr
& -\frac{i}{4}(\bar{\cd}^+_{\dt{\a}}\bar{\cd}^{\dt{\a}+}\bar{W})\cd^{--}
+\frac{i}{4}(\cd^{\a-}\cd^+_{\a}W) +\bar{W}W~,\cr} \eqno(4.27)$$
which is related to $\bo_{\rm cov}$ via the `bridge' transform. i.e. 
$$ \bo_{\rm cov}=e^{-\cv}\bo_{\rm cov.~analytic}e^{\cv}~.\eqno(4.28)$$
The particular form of the operator $\bo_{\rm cov}$ in the spacetime-constant 
gauge-invariant backround (4.15) reads ({\it cf.} ref.~\cite{ikz})
$$\eqalign{
\bo_{\rm cov.~const} = & \bo +\hat{Z}\hat{\bar{Z}}+\frac{1}{2}\x^{+-}
-\frac{i}{2}\x^{--}\left[ (\q^+\q^+)\hat{\bar{Z}}+ (\bar{\q}^+\bar{\q}^+)\hat{Z}
+(\theta^+)^2(\bar{\theta}^+)^2 e^2_0\x^{--}\right] \cr
& +\left[ \frac{i}{2}\x^{++}(\q^-\pa\bar{\q}^-)+\frac{i}{2}\x^{--}(\q^+\pa\bar{\q}^+)
-\x^{+-}(\q^+\pa\bar{\q}^-)+{\rm h.c.}\right] \cr
& +\left[ \frac{1}{2}\x^{++}(\q^-D^-) -\frac{1}{2}\x^{+-}(\q^+D^-)+{\rm h.c.}\right]
+\frac{1}{4}\x^{++}D^{--}~.\cr}\eqno(4.29)$$
The `bridge' itself is given by
$$\eqalign{
\cv_{\rm const}~=~& \frac{1}{2}\x^{--}(\q^+\q^-)(\bar{\q}^+)^2
+\frac{1}{4}\x^{++}(\q^-)^2(\bar{\q}^+\bar{\q}^-) \cr
~& -\frac{1}{4}\x^{+-}\left[ 2(\q^+\q^-)(\bar{\q}^+\bar{\q}^-)+(\q^+)^2(\bar{\q}^-)^2
\right] -{\rm h.c.} \cr}\eqno(4.30)$$

To get the non-physical hypermultiplet propagator, eq.~(4.23) has to be changed according to 
subsect.~4.1. 

The hypermultiplet propagator defined by eqs.~(4.23), (4.29) and (4.30) seems to be too
complicated, which prevented me from doing explicit perturbative calculations in HSS with the 
use of it now (this work is in progress). It is not even obvious to me whether the HSS methods  
remain to be technically superior in comparison to the conventional quantum perturbation theory 
in components (or in N=1 superfields) when N=2 supersymmetry is spontaneously broken and N=2
superfield propagators are manifestly $\q$-dependent. Nevertheless, the qualitative picture 
remains to be the same as in the previous subsect.~4.2: the kinetic term of the N=2 vector gauge
superfield is dynamically 
generated, with the induced (dimensionless and momentum-dependent) gauge coupling constant  
$$e^2_{\rm ind}(p)=e^2 +O(p^2/m^2)~.\eqno(4.31)$$
A momentum dependence of $e^2_{\rm ind}$ is calculable, while its low-energy value 
$e^2$ is non-vanishing, being a function of the dimensionless ratio $m^2/\x$. Indeed, the 
modification (4.29) of the box operator in the low-energy limit essentially amounts to a shift 
of the hypermultiplet mass, which is clearly {\it the same\/} for both (i.e. physical and 
non-physical) hypermultiplet propagators. 

The dynamical generation of the whole N=2 vector multiplet implies, of course, the dynamical 
deneration of all of its physical components, i.e. a complex scalar which can be interpreted
as a `Higgs' particle, a chiral spinor doublet representing a complex `photino', and a real 
`photon'. In the next sect.~5 we interpret the composite N=2 vector multiplet components as the
zero modes of a superstring ending on a {\it Dirichlet} (D) 6-brane. The non-perturbative 
phenomenon of the gauge symmetry enhancement for coincident D-6-branes appears to be surprisingly 
connected to the perturbative field theory considerations above via M-theory.

\section{Relation to M-theory and brane technology}

An exact solution to the LEEA of N=2 super-QCD with $N_c$ colors in spacetime $R^{1,3}$ can be 
identified with the LEEA of the effective (called N=2 MQCD) field theory defined in a single 
M-5-brane worldvolume given by the local product of $R^{1,3}$ and a hyperelliptic curve $\S_g$ of 
genus $g=N_c-1$ \cite{witten}.~\footnote{See e.g., refs.~\cite{ber,bt} for a review and 
ref.~\cite{polb} for an introduction.} The hyperelliptic curve $\S_g$ has to be holomorphically 
embedded into the hyper-K\"ahler 
four-dimensional multicentre Taub-NUT space $Q_{\rm mTN}$ associated with a multiple KK monopole. 
The identification of LEEA in these two apparently very different field theories (namely, the N=2 
super-QCD in the Coulomb branch, on the one side, and the N=2 MQCD defined in the M-5-brane 
worldvolume, on the other side) is highly non-trivial, since the former is defined as the leading 
contribution to the {\it quantum} LEEA in a gauge field theory, whereas the latter is determined by
{\it classical} M-5-brane dynamics or by eleven-dimensional (11d) supergravity equations of motion 
whose extended BPS solutions preserving some part of 11d supersymmetry are called M-theory branes.

\subsection{Multiple KK monopole}

The multiple KK monopole is a non-singular (solitonic) BPS solution to the classical equations of 
motion of 11d supergravity, with 11d spacetime being the product of the seven-dimensional 
(flat) Minkowski spacetime $R^{1,6}$ and the four-dimensional euclidean multicentre Taub-NUT 
space $Q_{\rm mTN}$ \cite{kkm1,kkm2,town}:
$$ ds^2_{[11]}  = dx^{m}dx^{n}\h_{mn}+ ds^2_{[4]}~, \qquad F_{(4)}\equiv dA_{(3)}=0~,
\eqno(5.1) $$
where $ds^2_{[4]}$ has been already defined in eq.~(2.11), $m=0,1,2,3,7,8,9$, $\vec{y}=\{y_i\}$, 
$i=4,5,6$. The eleventh coordinate in 11d has been identified with the periodic coordinate 
$\varrho$ of the multi-Taub-NUT space (2.11) with the harmonic function $H$ defined by eq.~(2.21).
The moduli $(k_A,\vec{y}_A)$ in eq.~(2.21) are interpreted as charges and locations of KK 
monopoles. For instance, the simple Taub-NUT space $(p=1)$ can be thought of as a 
non-trivial bundle (Hopf fibration) with the base $R^3$ and the fiber $S^1$ of magnetic charge $k$.
In general $(p\geq 1)$, there exist $p$ linearly independent normalizable self-dual harmonic 
2-forms $\o_A$ in $Q_{\rm mTN}$, which satisfy the orthogonality condition \cite{rub}
$$\fracmm{1}{(2\p k)^2}\int_{Q_{\rm mTN}} \,\o_A\wedge \o_B=\d_{AB}~.\eqno(5.2)$$

Two adjacent KK monopoles are connected by a homology 2-sphere having poles at the positions of the
monopoles. Near a singularity of $H$, the KK circle $S^1$ contracts to a point. A {\it holomorphic}
embedding of the Seiberg-Witten spectral curve $\S_g$ into the hyper-K\"ahler manifold 
$Q_{\rm mTN}$ is the consequence of the BPS condition \cite{mikh,heyi} 
$$ {\rm Area}\low{\,\S}=\abs{\int_{\S}\,\O_{\S} }~,\eqno(5.3)$$
where $\O_{\S}$ is the pullback of the K\"ahler $(1,1)$ form $\O$ of $Q_{\rm mTN}$ on $\S_g$. In
fact, any four-dimensional hyper-K\"ahler manifold possesses a holomorphic $(2,0)$ form $\o$, 
which is simply related to the K\"ahler form $\O$ as \cite{ath}
$$ \O^2=\o\wedge \bar{\o}~.\eqno(5.4)$$

The BPS states in M-theory, whose zero modes appear in the effective field theory defined in
the M-5-brane worldvolume, correspond to the supermembranes ( or M-2-branes) having minimal area
(BPS~!) and ending on the M-5-brane. The {\it spacial} topology of such M-2-brane determines the 
type of the corresponding N=2 supermultiplet in the effective (macroscopic) spacetime $R^{1,3}$: 
a cylinder $(Y)$ leads to an N=2 vector multiplet, whereas a disc $(D)$ gives rise to a 
hypermultiplet \cite{mikh,heyi}. Since the pullback $\o_Y$ on $Y$ is closed \cite{heyi}, there 
exists a meromorphic differential $\l_{SW}$ satisfying the relations $\o_Y=d\l_{SW}$ and
$$ Z=\int_{Y}\,\o_Y =\oint_{\pa Y}\l_{\rm SW}~,\eqno(5.5)$$
where $Z$ is the central charge and $\pa Y\in \S$. Hence, $\l_{\rm SW}$ can be identified with the 
Seiberg-Witten differential which determines the N=2 gauge LEEA in $R^{1,3}$ \cite{fsp}.

\subsection{Hypermultiplet LEEA from D-6-brane dynamics} 

As a result of KK compactification on the Seiberg-Witten curve $\S_g\,$, the leading 
{\it Nambu-Goto} (NG) term (proportional to the M-5-brane worldvolume) in the effective 
(six-dimensional) M-5-brane action reduces in the low-energy approximation to a four-dimensional 
scalar NLSM having the special K\"ahler geometry. This is enough to unambiguously restore the full
Seiberg-Witten LEEA \cite{sw} by $N=2$ supersymmetrization of the special bosonic NLSM: one
considers the NLSM complex scalars as the leading components of abelian N=2 vector multiplets in 
four spacetime dimensions, and then one deduces the Seiberg-Witten holomorphic potential $\cf$ out 
of the known special K\"ahler NLSM potential (see ref.~\cite{bt} for details)
$$ K(\F,\bar{\F})={\rm Im}\,\left(
\bar{\F}^i\fracmm{\pa\cf}{\pa\F^i}\right)~.\eqno(5.6)$$ 

Being applied to a derivation of the {\it hypermultiplet} LEEA of $N=2$ super-QCD in the 
Coulomb branch, brane technology suggests to dimensionally reduce the effective action of
a D-6-brane (to be described in M-theory by a KK-monopole) down to four spacetime 
dimensions \cite{bt}. In a static gauge for the D-6-brane, the indiced metric in the brane 
worldvolume is given by
$$ g\low{\tilde{\m}\tilde{\n}}=\h\low{\tilde{\m}\tilde{\n}}+G_{mn}\pa\low{\tilde{\m}}y^m
\pa\low{\tilde{\n}}y^n ~,\eqno(5.7)$$
where $\tilde{\m},\tilde{\n}=0,1,2,3,7,8,9$, $m,n=4,5,6,10$, and $G_{mn}$ is the multicentre 
euclidean Taub-NUT metric. After expanding the NG-part of the D-6-brane effective action  
$$ S_{\rm NG} =\int d^7\x\,\sqrt{-\det(g\low{\tilde{\m}\tilde{\n}})} \eqno(5.8)$$
up to the second order in the spacetime derivatives and performing a plain dimensional reduction
from seven to four spacetime dimensions, one arrives at the hyper-K\"ahler NLSM
$$ S[y]=\ha \int d^4x\,G_{mn}(y)\pa_{\m}y^m\pa^{\m}y^n~,\qquad
\m=0,1,2,3~,\eqno(5.9)$$
whose N=2 supersymmetrization yields the full N=2 supersymmetric hypermultiplet LEEA, in agreement
with the N=2 supersymmetric quantum field theory calculations in HSS \cite{ikz}.

\subsection{Symmetry enhancement for two coincident D-6-branes}

We are now in a position to discuss the symmetry enhancement in the case of two nearly coincident 
D-6-branes. The non-singular interpretation of D-6-branes in M-theory is based on the fact that the
isolated singularities of the harmonic function (2.21) are merely the {\it coordinate} 
singularities of the 
eleven-dimensional metric (5.1), though they are truly singular with respect to the dimensionally 
reduced ten-dimensional metric which is associated with D-6-branes in the type-IIA picture. The 
physical significance of the ten-dimensional metric singularities is now understood due to the 
illegitimate neglect of the KK modes related to the compactification circle $S^1$, since these KK 
particles (also called D-0-branes) become massless near the D-6-brane core \cite{town}. Their
inclusion is equivalent to accounting for instanton corrections in the four-dimensional N=2 
supersymmetric gauge field theory. 

When some parallel and similarly oriented D-branes coincide (this may happen in some special 
points of the moduli space of M-theory), it is accompanied by a gauge symmetry enhancement 
\cite{ht,witten2}. Since the brane singularities become non-isolated in the coincidence limit, 
first, they have to be resolved by considering the branes to be separated by some distance $\x$. 
Then one takes the limit of small $\x$. In the case of two parallel D-6-branes one substitutes the 
harmonic function (3.1), describing the double-centered Taub-NUT metric with a constant potential 
$\l$ at infinity, into eq.~(5.1). Then it describes two parallel and similarly oriented M-theory KK
monopoles with both centers on a line $\vec{\x}$ in the sixth direction, which dimensionally reduce
to a double D-6-brane configuration in ten dimensions. The homology 2-sphere connecting two KK 
monopoles contracts to a point in the limit $\x\to 0$, which gives rise to a curvature singularity 
of the dimensionally reduced metric in ten dimensions. From the eleven-dimensional perspective, 
M-2-branes can wrap about the 2-sphere connecting the KK monopoles, while the energy of the wrapped
M-2-brane is proportional to the area of the sphere \cite{ht}. When the sphere collapses, its area 
vanishes and, hence, the zero modes of the wrapped M-2-brane become massless, thus giving rise to
an extra massless vector supermultiplet in the LEEA and, hence, the gauge symmetry enhancement 
$$ U(1)\times U(1)~ \to~ U(2)\eqno(5.10)$$
assiciated with the $A_1$-type singularity. A non-perturbative phenomenon of the gauge symmetry 
enhancement was first observed in a K3 compactification of M-theory due to collapsing 2-cycles of K3
on the basis of duality with the heterotic string compactifications \cite{ov,kl}. It is worth 
mentioning here that the geometry near a collapsing 2-cycle of K3 is the same as the geometry near 
two almost coincident parallel KK monopoles in M-theory \cite{sen}, i.e. the corresponding 
hyper-K\"ahler metric is governed by the harmonic function (3.1) in the limit $\x\to 0$. 

From the ten-dimensional viewpoint, the wrapped M-2-branes are just the 6-6 superstrings stretched 
between two D-6-branes, so that it is the zero modes of these 6-6 superstrings that become massless
in the coincidence limit for D-6-branes (Fig.~3).

\begin{figure}
\vglue.1in
\makebox{
\epsfxsize=4in
\epsfbox{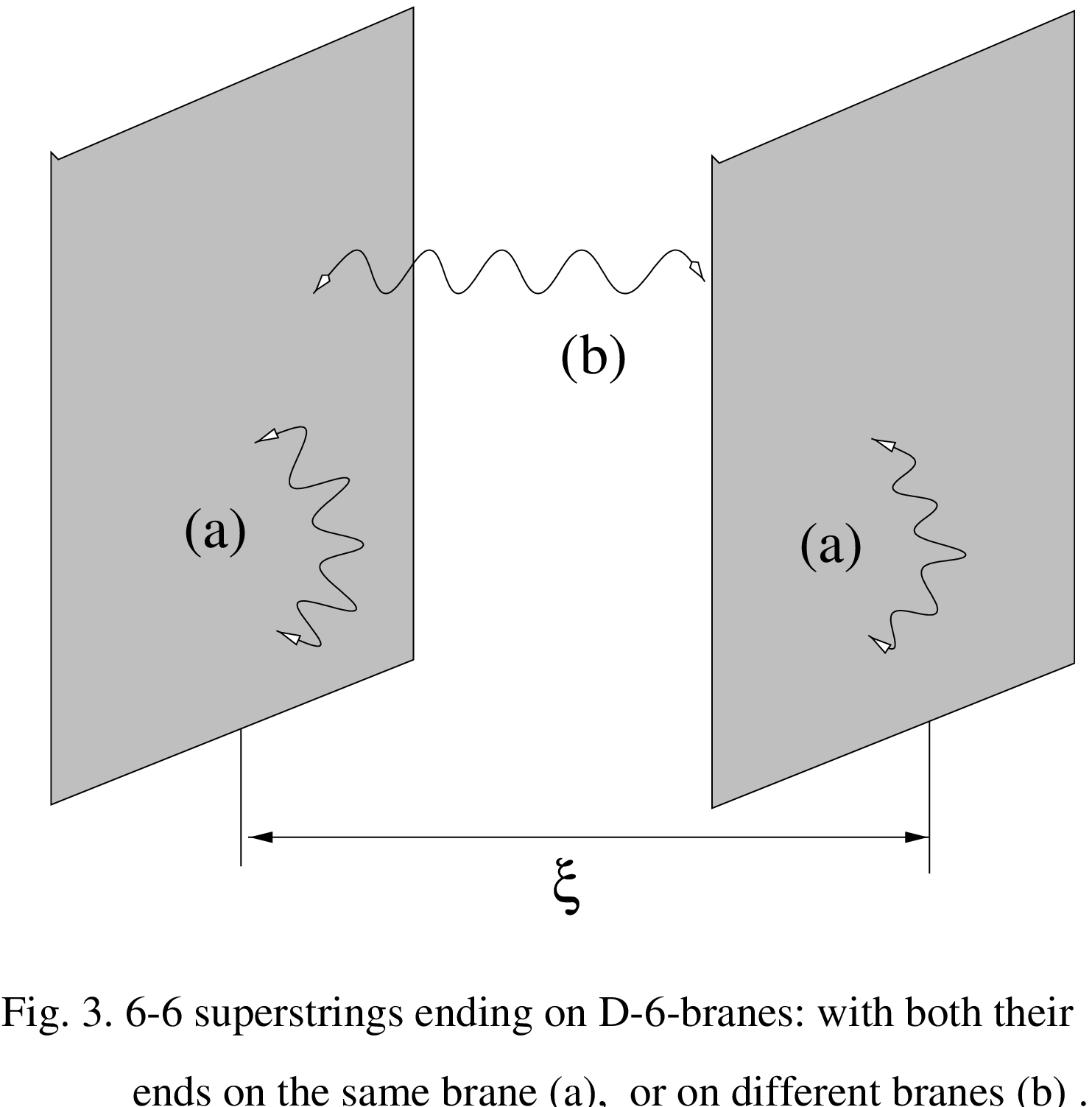}
}
\end{figure}

Each of the $U(1)$ factors on the left-hand-side of eq.~(5.10) is associated with a single
D-6-brane, being related to a 6-6 superstring whose both ends are on this brane (i.e. of type (a) 
in Fig.~3). The massless zero modes of this 6-6 superstring define an $U(1)$ gauge vector 
supermultiplet in the field theory LEEA describing the dynamics of small fluctuations about a 
D-6-brane. We can therefore identify this abelian vector supermultiplet with the composite vector 
supermultiplet dynamically generated from the hypermultiplet low-energy LEEA in the D-6-brane 
worldvolume (sect.~4).

Unlike the N=2 gauge LEEA, the {\it exact} hypermultiplet LEEA is entirely determined by its 
{\it one-loop} (perturbative) contribution having the form of the NLSM whose target space metric is
equal to the KK-monopole metric \cite{bt}. Unlike the 6-6 superstrings of the type (a) in Fig.~3, 
the 6-6 superstrings of the type (b) (see Fig.~3) with their ends on different D-6-branes cannot be
understood this way, being of truly non-perturbative origin. Indeed, our quantum field theory (3.8)
becomes singular in the limit $\x\to 0$. Accordingly, the {\it non-abelian} gauge symmetry 
enhancement (5.10) is beyond the scope of the hypermultiplet LEEA approach alone, which is 
apparently limited to a single D-brane worldvolume.
 
The hypermultiplet LEEA is obtained by a 4d spacetime N=2 supersymmetrization of the bosonic NLSM 
(5.9) whose hyper-K\"ahler metric (2.11) is fixed by the harmonic function (3.1). The abelian gauge
symmetry in terms of the hypermultiplet LEEA metric amounts to the gauged isometries in the 
corresponding NLSM target space, while their gauging itself can be made manifest in HSS. 
As was already mentioned in sect.~3, the N=2 supersymmetric NLSM with the double Taub-NUT metric
(two KK monopoles) is equivalent to the one with the {\it mixed} (Eguchi-Hanson-Taub-NUT) metric. 
The corresponding HSS action is given by eq.~(3.3) which makes it clear that the mixed 
hyper-K\"ahler NLSM metric does interpolate between the Eguchi-Hanson metric $(\l=0)$ and the 
Taub-NUT metric $(\x=0)$, both having the maximal isometry group $U(2)$. The action of the $U(2)$ 
isometry is linear in both limiting cases, while it is even holomorphic in the second case. Within 
the HSS approach this internal symmetry enhancement $U(1)\to SU(2)$ can be understood either as a 
restoration of the $SU(2)_A$ automorphism symmetry of N=2 supersymmetry algebra in the Taub-NUT 
limit, or as a restoration of the $SU(2)_{\rm PG}$ symmetry in the Eguchi-Hanson limit \cite{bt}.

On the one side, the geometry of two almost coinciding D-6-branes near the origin $\vec{y}=0$ can 
be approximated by the Eguchi-Hanson metric in M-theory since a finite asymptotical potential $\l$ 
in eq.~(3.1) becomes irrelevant near the singularity $\vec{y}=0$. The corresponding hypermultiplet 
LEEA (3.3) then reduces to our model (3.8) whose one-loop quantum fluctuations were investigated in
the preceeding sect.~4. On the other side, a D-6-brane naturally has in its worldvolume a massless 
abelian vector supermultiplet which can be understood as the Nambu-Goldstone mode associated with 
the 11d symmetries broken by the D-6-brane (BPS~!) classical solution \cite{bt}. Therefore, the 
dynamical generation of an abelian N=2 vector multiplet in the quantized 4d field theory (3.8) is 
consistent with the effective classical dynamics of two nearly coincident D-6-branes.
 
It may be natural to treat {\it all\/} 6-6 superstrings (i.e. of both types (a) and (b) in Fig.~3) 
on equal footing, like their zero modes. It is then tempting to conjecture that {\it any\/} 
(T-dual) open superstring ending on a D-brane should be considered as a {\it composite} (bound
state) of the D-brane physical degrees of freedom. 

\section{Conclusion and outlook}

The 11d supergravity approximation to M-theory is only valid for well-separated KK monopoles.
When KK monopoles coincide, their low-energy dynamics should be approximated by weakly coupled 
(perhaps, composite) superstrings propagating in the {\it multi-Eguchi-Hanson\/} (ALE) background 
\cite{sen}. The corresponding metric was found by Gibbons and Hawking in ref.~\cite{gheh}, while it
naturally originates as a particular limit of the multi-Taub-NUT (multicentre) metric, as we have 
already seen in the preceeding sections in the case of two coincident KK monopoles. 

When $p\geq 2$ ~D-6-branes coincide, i.e. all the moduli $\vec{y}_A$ 
in the harmonic function (2.21) are set to be zero, an additive 
asymptotic potential $\l$ can be always ignored near the core of these D-6-branes on top of each 
other. The multi-Eguchi-Hanson ALE space thus possesses an $A_{p-1}$ simple singularity which 
implies the enhanced non-abelian gauge symmetry $U(p)$ in the effective supersymmetric field 
theory defined in the common worldvolume of the coincident D-6-branes \cite{witten2}. Indeed, the 
effective gauge field theory is supposed to be defined in the limit where gravity decouples. 
The 11d supergravity has a 3-form $A^{[11]}_{(3)}$ which is decomposed with respect to the product 
of the D-6-brane worldvolume $R^{1,6}$ and the multi-Taub-NUT space $Q_{\rm mTN}$ as ({\it cf.\/}
ref.~\cite{kl})
$$ A^{[11]}_{(3)}=\sum_{B=1}^{p} A_{B(1)}^{[7]}\wedge \o_{B(2)}^{[4]} ~,\eqno(6.1)$$
where the 2-forms $\o_B$ in $Q_{\rm mTN}$ have been introduced in subsect.~5.1, whereas
$A_B$ are $p$ massless vectors (1-forms) in $R^{1,6}$. In addition, there are $3p$ scalar 
fields associated with the translational zero modes (or moduli) $\vec{y}_A$. Taken together, 
these vectors and scalars constitute the bosonic components of $p$ massless vector supermultiplets 
in $1+6$ dimensions, each having $8_{\rm B}+8_{\rm F}$ on-shell components. The gauge group of the 
effective field theory (in the case of separated KK monopoles) is therefore given by $U(1)^p$. 
Since the intersection matrix of 2-cycles in $Q_{\rm mTN}$ is known to be given by the Cartan 
matrix of $A_{p-1}$, the abelian gauge symmetry $U(1)^{p}$ should be enhanced to $U(p)$ in the 
coincidence limit. The area of the 2-cycles vanishes in this limit, so that the M-2-branes wrapped 
around these 2-cycles lead to the additional massless vectors to be identified with the M-2-brane 
zero modes. 

In the type-IIA picture, the 6-6 superstrings stretched between separated D-6-branes do not 
contribute to the effective LEEA in the Coulomb branch at all \cite{witten}. However, since the 
zero modes of these 6-6 superstrings become massless when the brane separation vanishes, they do 
contribute to the LEEA in our case, which may be called the {\it non-abelian} Coulomb branch. 
After a plain 
dimensional reduction from $R^{1,6}$ to $R^{1,3}$, the effective N=1 super-Yang-Mills theory in 
$1+6$ dimensions gives rise to the N=4 super-Yang-Mills theory in $1+3$ dimensions, which has the 
same number of on-shell physical components.  

A (spacetime) four-dimensional N=2 supersymmetric NLSM having the four-dimensional 
multi-Eguchi-Hanson target space can be constructed, for example, in HSS by coupling $p+1$ 
hypermultiplets in the fundamental 
representation of $SU(1,p)$ whose {\it Cartan subalgebra} (CSA) $u(1)^p$ is gauged by using $p$ 
abelian N=2 vector gauge superfields entering the NLSM action as Lagrange multipliers in the 
presence of FI terms for all of them ({\it cf.\/} eq.~(3.8) and ref.~\cite{giot}). There is, 
however, a problem with this approach because of the mismatch between the numbers of physical and 
non-physical hypermultiplets (unless $p=1$), which may lead to UV divergences in 4d and, hence, 
quantum inconsistencies. A possible resolution may be just taking into account more physical 
hypermultiplets. Above we only considered a single physical hypermultiplet whose LEEA (NLSM) metric
had a direct geometrical interpretation in M-theory in terms of four-dimensional (KK) monopoles. 
However, more physical hypermultiplets can appear in the field theory LEEA, e.g., after taking into 
account the zero modes of the 4-6 superstrings stretched between D-4- and D-6-branes in N=2 MQCD 
\cite{witten}.~\footnote{The M-5-brane discussed in the beginning of 
sect.~5 represents in 11d M-theory an intersecting \newline ${~~~~~}$ configuration of two NS 
5-branes and $N_c$ D-4-branes in ten dimensions (=type-IIA picture).} The hypermultiplet LEEA, 
capable to dynamically generate a non-abelian N=2 vector gauge multiplet, might be 
given by the N=2 supersymmetric gauged NLSM over the non-compact coset $SU(p,p)/U(p)$ in HSS, whose
$p$ physical and $p$ non-physical hypermultiplets together are in some (other than fundamental) 
representation of $SU(p,p)$, with the $U(p)$ subgroup being gauged by a non-abelian N=2 vector 
gauge superfield. However, there is another caveat here since FI terms only exist for abelian gauge
groups. It is not clear to me how to formulate a corresponding N=2 supersymmetric NLSM action, 
if any.

A unique consistent possibility seems to be the hypermultiplet LEEA having the form of a gauged N=2
NLSM over the coset $SU(p,p)/U(1)^p$. The hypermultiplets then belong to the fundamental 
representation of $SU(p,p)$ whose abelian subgroup $U(1)^p$ is to be gauged. The latter is supposed
to be associated with the CSA of the full non-abelian gauge group $U(p)$ which only appears
after taking into account non-perturbative corrections due to D-6-branes. 

The starting N=2 NLSM action in the analytic HSS is given by  
$$ S[q,V] = \int_{\rm analytic}\left\{ \tr\low{\,\rm fund}
\left(\sbar{q}{}^{+}\cd^{++} q^+\right)+\tr\low{\,\rm CSA}\left(V^{++}\,\x^{++} \right) 
\right\}~,\eqno(6.2)$$
where the $u(1)^p$ CSA-valued harmonic- and gauge-covariant derivative with central charge, 
$\cd^{++}=D_{\hat{Z}}^{++}+iV^{++}$, has been introduced. The FI terms in eq.~(6.2) are also 
necessary in order to get rid of a singularity which would appear in their absence.

It is straightforward to calculate the local part of the one-loop effective action 
$i\Tr\log\cd^{++}$ along the lines of sect.~4 in the LEEA approximation, with the latter being 
defined by the condition $p^2\ll\x^2$ for all external momenta $p^{\m}$. It should result in a 
dynamical generation of an N=2 supersymmetric $U(1)^p$ gauge field theory, whose unique N=2 
supersymmetric non-abelian $U(p)$ extension is given by a standard N=2 {\it super-Yang-Mills} 
(SYM) action. The latter reads in HSS as follows \cite{zup}:
$$\eqalign{
S_{\rm N=2~SYM}[V] ~=~ &  \fracmm{1}{g^2\low{\rm YM}}\int d^4xd^4\q d^4\bar{\q}\, 
\tr \sum^{\infty}_{n=2}
\fracmm{(-i)^n}{n}\int du_1du_2\cdots du_n\times  \cr
~ & \times \fracmm{ V^{++}(Z,u_1)V^{++}(Z,u_2)\cdots V^{++}(Z,u_n)}{(u^+_1u^+_2)
(u^+_2u^+_3)\cdots (u^+_nu^+_1)}~.\cr}\eqno(6.3)$$

The induced `running' gauge coupling constant takes the form ({\it cf.} sect.~4)
$$ \fracmm{1}{g^2\low{\rm YM}}= \fracmm{C_A}{16\p^2} \int^1_0 dx\,\ln \fracmm{ \abs{Z}^2+\x
+p^2x(1-x)}{\abs{Z}^2+p^2x(1-x)}~,\eqno(6.4)$$
where we introduced the gauge group generators $t_a$ in the fundamental representaton of $U(p)$, 
subject to $\tr(t_at_b)=C_A\d_{ab}$, the physical central charge $Z$ and 
$\x=\sqrt{(\vec{\x})^2}>0$. 
 
A different gauge symmetry enhancement pattern appears when $p$ of D-6-branes come on top of an 
{\it orientifold} six-plane, which leads to the $SO(2p)$ gauge symmetry \cite{ov}. The orientifold 
six-plane can be represented in M-theory by the (hyper-K\"ahler) {\it Atiyah-Hitchin} space 
\cite{ath} instead of a KK monopole. Indeed, far from the origin the Atiyah-Hitchin space has the 
topology $R^3\times S^1/\ct_4$, i.e. it looks like $Q_{\rm mTN}$ whose points are now supposed to 
be identified under the action of the discrete symmetry $\ct_4$ reversing signs of all four 
coordinates of $Q_{\rm mTN}$. This matches the definition of the orientifold six-plane according to
ref.~\cite{sen}. It is now straightforward to generalize our discussion to orthogonal gauge groups
too.

To this end, let's briefly consider the conformally invariant limit of the hypermultiplet LEEA
corresponding to the case of $N_c$ coincident D-6-branes $(\x\to 0)$ in N=2 MQCD at large number 
of colors $N_c$. The physical hypermultipets in the 4d effective (N=2 supersymmetric, by 
construction) field theory then form the {\it adjoint} representation of the gauge group, so that 
we arrive at a sum of the gauge-invariant hypermultiplet LEEA and the induced N=2 SYM action (6.3),
$$
\int_{\rm analytic} \tr\low{\rm ad}\left(\sbar{q}{}^{+}\cd^{++} q^+\right)+ S_{\rm N=2~SYM}
\equiv S_{\rm N=4~SYM}~,\eqno(6.5)$$
which is `almost' the N=4 supersymmetric Yang-Mills action in the 4d, N=2 harmonic superspace, whose
N=4 supersymmetry is merely broken down to N=2 by a non-vanishing central charge $Z$. The induced 
N=4 SYM coupling constant in the low-energy limit $(p^2=0)$ but at large number $N_c$ of colors 
(i.e. in the t'Hooft limit) satisfies the relation
$$ N_cg^{2}_{\rm YM}\sim \fracmm{\abs{Z}^2}{\abs{\x}}~.\eqno(6.6)$$
It should be compared to the recent conjecture of Maldacena \cite{mal}. He discussed a classical 
BPS 
solution describing a `plain' M-5-brane in the particular limit (down the `throat') given by the 
product $AdS_7\times S^4$ whose both radii are proportional to $N_c^{1/3}$. For large $N_c$ the 
Maldacena LEEA is given by a $(2,0)$ superconformally invariant gauge field theory in six 
dimensions, which is supposed to be dual to M-theory compactified on $AdS_7\times S^4$ \cite{mal}. 
Down to four spacetime dimensions, Maldacena considered $N_c$ D-3-branes at large $N_c$ instead, 
and he argued that the N=4 SYM theory in the t'Hooft limit has to be dual to the IIB superstring 
theory compactified on $AdS_5\times S^5$. The four-dimensional N=4 SYM theory is defined on the 
boundary of the $AdS_5$-space, with the correspondence
$$ N_cg^{2}_{\rm YM}\sim (\a')^{-2}R^4_{\rm AdS} \quad {\rm and}\quad
g^2\low{\rm YM}\sim g\low{\rm string} ~.
\eqno(6.7)$$

Note that the t'Hooft limit of large $N_cg^{2}_{\rm YM}$ is equivalent to $\abs{\x/Z^2}\to 0$ in our
approach. Hence, the conformal (t'Hooft-Maldacena) LEEA limit described by the `almost' N=4 
supersymmetric Yang-Mills theory can be deduced from the hypermultiplet LEEA near the singularity,
after taking into account a non-vanishing physical central charge, a dynamical generation of the 
abelian N=2 vector gauge multiplets associated with the SCA of the gauge group, and a 
non-perturbative non-abelian gauge symmetry enhancement. A possible dynamical orgin of the N=2
central charge is left as an open problem.

\vglue.2in
 
\section*{Acknowledgements}

I am grateful to Joseph Buchbinder, Evgeny Ivanov, Olaf Lechtenfeld, J\"urgen Schulze, Kelly Stelle
and Boris Zupnik for stimulating discussions. I also wish to thank the Bogoliubov Laboratory of 
Theoretical Physics at Joint Institute for Nuclear Research, where a part of this work was done, 
for a kind hospitality extended to me in Dubna. 

\newpage

\section*{Appendix A: N=2 harmonic superspace (HSS)}

Four-dimensional field theories with N=2 extended supersymmetry can be formulated in the 
conventional N=2 extended superspace parameterized by the coordinates 
$Z^M=(x^{\m},\q^{\a}_i,\bar{\q}^{\dt{\a}i})$, $\m=0,1,2,3$, $\a=1,2$, $i=1,2$, and 
$\Bar{\q^{\a}_i}=\bar{\q}^{\dt{\a}i}$, in terms of constrained N=2 superfields \cite{oldrev}. 
Unfortunately, the standard constraints \cite{fs,gs},
$$ \{ \cd^i_{\a},\bar{\cd}_{\dt{\a}j}\}=-2i\d^i_j\cd_{\a\dt{\a}}~,\quad
\{\cd^i_{\a},\cd^j_{\b}\}=-2\ve_{\a\b}\ve^{ij}\bar{W}~,\quad 
\{ \bar{\cd}_{\dt{\a}i},\bar{\cd}_{\dt{\b}j}\}=-2\ve_{\dt{\a}\dt{\b}}\ve\low{ij}W~,
\eqno(A.1)$$
and 
$$ \cd_{\a}{}^{(i}q^{j)}=\bar{\cd}_{\dt{\a}}{}^{(i}q^{j)}=0~,\eqno(A.2)$$
defining a (non-abelian) N=2 vector multiplet and a Fayet-Sohnius hypermultiplet, respectively,
in the conventional N=2 superspace in terms of the gauge- and super-covariant 
(Lie algebra-valued) N=2 superspace derivatives
$$ \cd_M\equiv (\cd_{\m},\cd^i_{\a},\bar{\cd}_{\dt{\a}i})=D_M+\ca_M~,\eqno(A.3)$$
do not have a manifestly holomorphic (or analytic) structure. Accordingly, they also do not have
a simple solution in terms of {\it unconstrained} N=2 superfields which are needed for a 
manifestly supersymmetric quantization. The situation is even more dramatic for the FS 
hypermultiplet, since its defining equations (A.2) (in the absence of central charges) are
merely on-shell constraints whereas the known off-shell formulations of a hypermultiplet 
in the conventional $N=2$ superspace are either not universal (like an N=2 tensor multiplet 
\cite{klr} ) or very cumbersome (like a relaxed hypermultiplet \cite{hst} or the generalized 
N=2 tensor multiplets \cite{lkt}) so that their practical meaning is limited \cite{nlsmbook}.

In the HSS formalism, the standard N=2 superspace is extended by adding bosonic variables 
(or `zweibeins') $u^{\pm i}$ parameterizing a 2-sphere $S^2\sim SU(2)/U(1)$. By using these
extra variables one can make manifest the hidden analyticity structure of the standard N=2 
superspace constraints (A.1) and (A.2) as well as to find their manifestly
N=2 supersymmetric solutions in terms of unconstrained (analytic) N=2 superfields. The harmonic
variables have the property
$$ \left( \begin{array}{c} u^{+i} \\ u^{-i}\end{array}\right) \in SU(2)~,\eqno(A.4)$$
so that
$$ u^{+i}u^-_i=1~,\quad u^{+i}u^+_i=u^{-i}u^-_i=0~,\quad{\rm and}\quad \Bar{u^{i+}}=u^-_i~.
\eqno(A.5)$$
Instead of using an explicit parameterization of the sphere $S^2$,
it is convenient to deal with functions of zweibeins, that carry a definite
$U(1)$ charge $U$ to be defined by $U(u^{\pm}_i)=\pm 1$, and use the following
integration rules~\cite{gikos}:
$$ \int du =1~,\qquad \int du\, u^{+(i_1}\cdots u^{+i_m}u^{-j_1}\cdots
u^{-j_n)}=0~,\quad {\rm when}\quad m+n>0~.\eqno(A.6)$$
It is obvious that any integral over a $U(1)$-charged quantity vanishes.

The usual complex conjugation does not preserve analyticity. However, being combined with 
another (star) conjugation that only acts on $U(1)$ indices as 
$(u^+_i)^*=u^-_i$ and $(u^-_i)^*=-u^+_i$, it does preserve analyticity. One easily 
finds~\cite{gikos}
$$ \sbar{u^{\pm i}}=-u^{\pm}_i~,\qquad  \sbar{u^{\pm}_i}=u^{\pm i}~.
\eqno(A.7)$$

The covariant derivatives with respect to the zweibeins, preserving the
defining equations (A.4) and (A.5), are given by
$$ D^{++}=u^{+i}\fracmm{\pa}{\pa u^{-i}}~,\quad
D^{--}=u^{-i}\fracmm{\pa}{\pa u^{+i}}~,\quad
D^{0}=u^{+i}\fracmm{\pa}{\pa u^{+i}}-u^{-i}\fracmm{\pa}{\pa u^{-i}}~.
\eqno(A.8)$$
It is easy to check that they satisfy the $SU(2)$ algebra,
$$\[ D^{++},D^{--}\]=D^0~,\quad \[D^0,D^{\pm\pm}\]=\pm 2D^{\pm\pm}~,
\eqno(A.9)$$
and commute with the N=2 superspace derivatives (A.3). Eq.~(A.9) is supposed to be added 
to the constraints (A.1) and (A.2).

The key feature of the N=2 HSS is the existence of the so-called
 {\it analytic} subspace parameterized by the coordinates
$$ (\z,u)=\left\{ \begin{array}{c}
x^{\m}_{\rm A}=x^{\m}-2i\q^{(i}\s^{\m}\bar{\q}^{j)}u^+_iu^-_j~,~~
\q^+_{\a}=\q^i_{\a}u^+_i~,~~ \bar{\q}^+_{\dt{\a}}=\bar{\q}^i_{\dt{\a}}u^+_i~,~~
u^{\pm}_i \end{array} \right\}~,\eqno(A.10)$$
which is invariant under N=2 supersymmetry and closed under the combined
conjugation of eq.~(A.7)~\cite{gikos}. This allows one to define {\it analytic}
superfields of any non-negative and integer $U(1)$ charge $q$ by the analyticity
conditions
$$D^+_{\a}\f^{(q)}=\bar{D}^+_{\dt{\a}}\f^{(q)}=0~,\quad {\rm where}\quad
D^{+}\low{\a}=D^i_{\a}u^+_i \quad {\rm and}\quad
\bar{D}^+_{\dt{\a}}=\bar{D}^i_{\dt{\a}}u^+_i~.\eqno(A.11)$$

The analytic measure reads $d\z^{(-4)}du\equiv d^4x_{\rm A}
d^2\q^+d^2\bar{\q}^+du$. It carries the $U(1)$ charge $(-4)$,
whereas the full neutral measure of $N=2$ HSS is given by
$$ d^4xd^4\q d^4\bar{\q}du=d\z^{(-4)}du(D^+)^4~,\eqno(A.12)$$
where
$$(D^+)^4=(D^+)^2(\bar{D}^+)^2
=\fracmm{1}{16}(D^{+\a}D_{\a}^+)(\bar{D}^{+}_{\dt{\a}}\bar{D}^{+\dt{\a}})~.
\eqno(A.13)$$
In the analytic subspace, the harmonic derivative $D^{++}$ reads
$$D^{++}_{analytic} = D^{++}-2i\q^+\s^{\m}\bar{\q}^+\pa_{\m}~,\eqno(A.14)$$
it preserves analyticity, and it allows one to integrate by parts. Both the
original (central) basis and the analytic one can be used on equal footing in
the HSS. In the main text and in what follows we omit the subscript {\it analytic}  at the
covariant derivatives in the analytic basis, in order to simplify our notation.

It is the advantage of the analytic N=2 HSS compared to the ordinary N=2
superspace that both an off-shell N=2 vector multiplet and an off-shell
hypermultiplet can be introduced there on equal footing. There exist two
off-shell hypermultiplet versions in HSS, which are dual to each other.
The so-called {\it Fayet-Sohnius-type} (FS) hypermultiplet is defined as an
unconstrained complex analytic superfield $q^+$ of $U(1)$-charge $(+1)$,
whereas its dual, called the {\it Howe-Stelle-Townsend-type} (HST) hypermultiplet,
is a real unconstrained analytic superfield $\o$ with the vanishing
$U(1)$-charge.~\footnote{It is worth mentioning here that both FS and HST multiplets
were originally introduced in the \newline ${~~~~~}$ {\it conventional} N=2
superspace~\cite{fs,hst}, whereas we use the same names to denote N=2 {\it harmonic} 
\newline ${~~~~~}$ superfields, which are different off-shell but reduce to the FS and HST 
multiplets on-shell.}
The on-shell physical components of the FS hypermultiplet comprise an $SU(2)$
doublet of complex scalars and a Dirac spinor which is a singlet with respect to 
$SU(2)$. The on-shell physical components of the HST hypermultiplet comprise a
real singlet and a real triplet of scalars, and a doublet of chiral spinors. The FS
hypermultiplets are natural for describing a charged N=2 matter, whereas the HST 
hypermultiplets are more appropriate for describing a neutral N=2 matter.

Similarly, an N=2 vector multiplet is described by an unconstrained analytic
superfield $V^{++}$ of the $U(1)$-charge $(+2)$. The $V^{++}$ is real in the
sense $\Bar{V^{++}}^{\,*}=V^{++}$, and it can be naturally introduced as a
connection to the harmonic derivative $D^{++}$.

A free FS hypermultiplet HSS action is given by (in canonical normalization) 
$$ S[q]=-\int d\z^{(-4)}du\,\sbar{q}{}^+D^{++}q^+~,\eqno(A.15)$$
whereas its minimal coupling to an abelian N=2 gauge superfield reads
$$ S[q,V]= - \int d\z^{(-4)}du \,\sbar{q}{}^+(D^{++}+iV^{++})q^+~.
\eqno(A.16)$$
It is not difficult to check, for example, that the free FS hypermultiplet equations of
motion, $D^{++}q^+=0$, imply $q^+=q^i(Z)u^+_i$ {\it and\/} the (on-shell) Fayet-Sohnius 
constraints (A.2) in the conventional N=2 superspace, 
$$D_{\a}^{(i}q^{j)}(Z)=D_{\dt{\a}}^{(i}q^{j)}(Z)=0~.\eqno(A.17)$$

Similarly, a free HSS action of the HST hypermultiplet is given by
$$S[\o]=-\frac{1}{2}\int d\z^{(-4)}du \,(D^{++}\o)^2~,\eqno(A.18)$$
and it is equivalent (dual) to the standard N=2 tensor (linear)
multiplet action \cite{nlsmbook}. 

The constraints (A.1) defining the N=2 super-Yang-Mills theory in the conventional N=2 
superspace imply the existence of a (covariantly) chiral~\footnote{A covariantly-chiral 
superfield can be transformed into a chiral superfield by field redefinition.} and 
gauge-covariant N=2 SYM field strength $W$ satisfying the reality condition 
(or the Bianchi `identity')
$$ {\cal D}^{\a}\low{(i}{\cal D}\low{j)\a}W=\bar{\cal D}_{\dt{\a}(i}
\bar{\cal D}^{\dt{\a}}\low{j)}\bar{W}~,\eqno(A.19)$$
which is a consequence of eq.~(A.1).

An N=2 supersymmetric solution to the non-abelian N=2 SYM constraints (A.1) in the ordinary 
N=2 superspace is not known in an analytic form (see, however, ref.~\cite{koll} for partial
results). It is the N=2 HSS reformulation of 
the N=2 SYM theory that makes it possible. An exact non-abelian relation between 
the constrained, harmonic-independent superfield strength $W$ and the unconstrained analytic 
(harmonic-dependent) superfield $V^{++}$ is given in refs.~\cite{gikos}, and it is highly 
non-linear and complicated. The abelian relation is simple, and it is given by
$$ W=\{ \bar{\cal D}^+_{\dt{\a}},\bar{\cal D}^{-\dt{\a}}\}
=-(\bar{D}^+)^2A^{--}~,\eqno(A.20)$$
where the non-analytic harmonic superfield connection $A^{--}(Z,u)$ to the
derivative $D^{--}$ has been introduced, ${\cal D}^{--}=D^{--}+iA^{--}$.

As a consequence of the N=2 HSS abelian constraint
$\[{\cal D}^{++},{\cal D}^{--}\]={\cal D}^0=D^0$, the connection $A^{--}$
satisfies the relation
$$ D^{++}A^{--}=D^{--}V^{++}~,\eqno(A.21)$$
whereas eq.~(A.19) can be rewritten to the form
$$ (D^+)^2W=(\bar{D}^+)^2\bar{W}~.\eqno(A.22)$$

A solution to the $A^{--}$ in terms of the analytic unconstrained superfield
$V^{++}$ easily follows from eq.~(A.21) when using the identity~\cite{gikos}
$$ D^{++}_1(u_1^+u^+_2)^{-2}=D_1^{--}\d^{(2,-2)}(u_1,u_2)~,\eqno(A.23)$$
where we have introduced the harmonic delta-function $\d^{(2,-2)}(u_1,u_2)$
and the harmonic distribution $(u_1^+u^+_2)^{-2}$ according to their
definitions in ref.~\cite{gikos}, hopefully, in the self-explaining
notation. One finds~\cite{zupnik}
$$ A^{--}(z,u)= \int dv \,\fracmm{V^{++}(z,v)}{(u^+v^+)^2}~,\eqno(A.24)$$
and
$$ W(z)=-\int du (\bar{D}^-)^2V^{++}(z,u)~,\quad \bar{W}(z)=
-\int du (D^-)^2V^{++}(z,u)~,\eqno(A.25)$$
by using an identity
$$u^+_i=v^+_i(v^-u^+)-v^-_i(u^+v^+)~,\eqno(A.26)$$
which is the obvious consequence of the definitions (A.5).

The free equations of motion of an N=2 vector multiplet are given by the vanishing analytic
superfield
$$ (D^+)^4A^{--}(Z,u)=0~,\eqno(A.27)$$
while the corresponding action reads~\cite{zupnik}
$$ \eqalign{
S[V]= & -\fracmm{1}{2e^2}\int d^4xd^4\theta\, W^2 =
-\fracmm{1}{2e^2}\int d^4xd^4\theta d^4\bar{\theta}du \,V^{++}(Z,u)A^{--}(Z,u)\cr
= & -\fracmm{1}{2e^2}\int d^4xd^4\theta d^4\bar{\theta}du_1du_2\,
\fracmm{V^{++}(Z,u_1)V^{++}(Z,u_2)}{(u_1^+u^+_2)^2}~,\cr}\eqno(A.28)$$
where we have introduced an electromagnetic coupling constant $e$.

In a WZ-like gauge, the abelian analytic HSS prepotential $V^{++}$ amounts
to the following explicit expression~\cite{gikos}:
$$\eqalign{
 V^{++}(x_{\rm A},\theta^+,\bar{\theta}^+,u)=&
\bar{\theta}^+\bar{\theta}^+a(x_{\rm A})
+ \bar{a}(x_{\rm A})\theta^+\theta^+
-2i\theta^+\s^{\m}\bar{\theta}^+V_{\m}(x_{\rm A}) \cr
& +\bar{\theta}^+\bar{\theta}^+\theta^{\a +}\j^i_{\a}(x_{\rm A})u^-_i
+\theta^+\theta^+\bar{\theta}^+_{\dt{\a}}\bar{\j}^{\dt{\a}i}(x_{\rm A})u^-_i\cr
&+\theta^+\theta^+\bar{\theta}^+\bar{\theta}^+ D^{(ij)}(x_{\rm A})u^-_iu^-_j~,
\cr} \eqno(A.29)$$
where $(a,\j^i_{\a},V_{\m},D^{ij})$ are the usual N=2 vector multiplet components 
(see Appendix B for more details).

A hypermultiplet (BPS) mass can only come from the central charges in the N=2 supersymmetry 
algebra since, otherwise, the number of the massive hypermultiplet
components has to be increased. The most natural way to introduce central
charges $(Z,\bar{Z})$ is to identify them with spontaneously broken $U(1)$
generators of dimensional reduction from six dimensions via the Scherk-Schwarz
mechanism~\cite{ikz}. Indeed, after being written down in six dimensions,
eq.~(A.14) implies an additional `connection' term in the associated
four-dimensional harmonic derivative,
$$ {\cal D}^{++}=D^{++}+v^{++}~,\quad {\rm where}\quad
v^{++}=i(\theta^+\theta^+)\bar{Z}+i(\bar{\theta}^+\bar{\theta}^+)Z~.
\eqno(A.30)$$
Comparing eq.~(A.30) with eqs.~(A.16), (A.25) and (A.29) clearly shows that the N=2
central charges can be equivalently treated as the abelian N=2 vector superfield
background with the covariantly constant chiral superfield strength \cite{bbiko}
$$ \VEV{W}=\VEV{a}=Z~.\eqno(A.31)$$

It is also worth mentioning here that introducing central charges into the algebra (A.1) of the 
N=2 superspace covariant derivatives implies corresponding
charges in a similar N=2 supersymmetry algebra and, hence, in the N=2 supersymmetry 
transformation laws of N=2 superfields and their components. The HSS formalism 
automatically incorporates these changes via simple modifications of the HSS covariant 
derivatives. Non-vanishing N=2 central charges also break the rigid R-symmetry
$$ \q^i_{\a}\to e^{-i\g} \q^i_{\a}~,\quad \bar{\q}^{\dt{\a}i}\to  e^{+i\g} \bar{\q}^{\dt{\a}i}~,
\eqno(A.32)$$
of a massless N=2 supersymmetric field theory. This fact alone is responsible for a presence of
anomalous (holomorphic) terms in the N=2 gauge low-energy effective action \cite{div,seib}.

\section*{Appendix B: N=2 restricted chiral superfield}

In this Appendix we collect the known facts about components of the restricted  N=2 chiral
superfield \cite{oldrev} describing the abelian N=2 vector gauge superfield strength $W$ in the
main text. It simultaneously defines the rest of our notation.

The convenient realization of the supercovariant derivatives in the ordinary N=2 superspace 
(with vanishing central charge) is given by
$$ D^i_{\a}=\fracmm{\pa}{\pa\q^{\a}_i} +i\bar{\q}^{\dt{\a}i}\pa_{\a\dt{\a}}~,\quad
\bar{D}_{\dt{\a}i}= -\fracmm{\pa}{\pa\bar{\q}^{\dt{\a}i}} -i\q^{\a}_i\pa_{\a\dt{\a}}~,
\eqno(B.1)$$
where we have used the standard two-component spinor notation, 
$\pa_{\a\dt{\a}}=\s^{\m}_{\a\dt{\a}}\pa_{\m}$~.

The restricted N=2 chiral superfield $W$ is an off-shell irreducible N=2 superfield satisfying
the constraints
$$ \bar{D}_{\dt{\a}i}W=0~,\qquad D^4W=\bo \bar{W}~.\eqno(B.2)$$
The first constraint of eq.~(B.2) is an N=2 generalization of the usual N=1 chirality condition,
whereas the second one can be considered as a generalized {\it reality} condition \cite{oldrev} 
having no analogue in N=1 superspace. A solution to eq.~(B.2) in N=2 chiral superspace 
$(y^{\m},\q_{\a}^i)$ reads
$$\eqalign{
W(y,\q)~=~& a(y) + \q^{\a}_i\j^i_{\a}(y)-\ha\q^{\a}_i\vec{\t}\ud{i}{j}\q^j_{\a}\cdot\vec{D}(y)\cr
~&  +\frac{i}{8}\q_i^{\a}(\s^{\m\n})\du{\a}{\b}\q_{\b}^iF_{\m\n}(y)-i(\q^3)^{i\a}\pa_{\a\dt{\b}}
\bar{\j}_i^{\dt{b}}(y)+\q^4\bo \bar{a}(y)~,\cr}\eqno(B.3)$$
where we have introduced a complex scalar $a$, a chiral spinor doublet $\j$, a real isovector
$\vec{D}=\ha(\vec{\t})\ud{i}{j}D\ud{j}{i}\equiv \ha\tr(\vec{\t}D)$, $\tr(\t_m\t_n)=2\d_{mn}$,
and a real antisymmetric tensor $F_{\m\n}$ as the field components of $W$, while $F_{\m\n}$ 
has to satisfy a spacetime constraint \cite{oldrev}
$$ \ve^{\m\n\l\r}\pa_{\n}F_{\l\r}=0 ~.\eqno(B.4)$$
Eq.~(B.4) can be interpreted as the `Bianchi identity' whose solution is given by
$$ F_{\m\n}=\pa_{\m}V_{\n}-\pa_{\n}V_{\m} \eqno(B.5)$$
in terms of a vector gauge field $V_{\m}$ subject to gauge transformations  
$\d V_{\m}=\pa_{\m}\l$.

N=2 supersymmetry transformation laws for the components are easily obtained by imposing the
scalar transformation law on the N=2 restricted chiral superfield (B.3) under chiral N=2 
supertranslations 
$$ \d y^{\m}=-i(\q_i\s^{\m}\bar{\ve}^i)~,\quad \d\q^{\a}_i=\ve^{\a}_i~.\eqno(B.6)$$
One easily finds \cite{oldrev}
$$\eqalign{
\d a~=~& \ve^{\a}_i\j^i_{\a}~,\cr
\d\j^i_{\a}~=~&-\vec{\t}\ud{i}{j}\cdot\vec{D}\ve^j_{\a}
-i\pa_{\a\dt{\b}}a\bar{\ve}^{\dt{\b}i}+
\frac{i}{4} (\s^{\m\n}\ve^i)_{\a} F_{\m\n}~,\cr
\d \vec{D}~=~& -\frac{i}{2} (\bar{\ve}_{\dt{\a}i}\vec{\t}\ud{i}{j}\pa^{\dt{\a}\a}\j^j_{\a}) 
+ {\rm h.c.}~,\cr
\d F_{\m\n}~=~ & -\pa_{\m}(\ve_i^{\a}\s_{\n\a\dt{\a}}\bar{\j}^{\dt{\a}i}) 
+\pa_{\n}(\ve_i^{\a}\s_{\m\a\dt{\a}}\bar{\j}^{i\dt{\a}})
+ {\rm h.c.} \cr}
\eqno(B.7)$$

A free N=2 supersymmetric Maxwell Lagrangian is given by
$$-\fracmm{1}{2e^2} \int d^4\q\, W^2= \fracmm{1}{e^2}\abs{\pa_{\m}a}^2 +
\fracmm{i}{4e^2}\bar{\j}_{\dt{\a}j}\pa^{\dt{\a}\a}\j^j_{\a} -\fracmm{1}{4e^2}F^2_{\m\n}
+ \fracmm{1}{2e^2}\vec{D}^2~~.\eqno(B.8)$$

Given the constraints (B.2), it is not difficult to verify that the N=2 superfield 
$$L^{ij}\equiv D^{ij}W\eqno(B.9)$$ 
satisfies \cite{oldrev}
$$ D_{\a}^{(i}L^{jk)}= \bar{D}_{\dt{\a}}{}^{(i}L^{jk)}=0~,\eqno(B.10)$$
and it is subject (up to a constant -- see below) to the reality condition 
$$(L^{ij})^{\dg}=\ve_{ik}\ve_{jl}L^{kl}~,\quad {\rm or,~equivalently,}\quad
\vec{L}\,{}^{\dg}=\vec{L}~.\eqno(B.11)$$

Eqs.~(B.10) and (B.11) are known as the defining constraints of an N=2 {\it tensor}
multiplet superfield $L^{ij}$ \cite{klr,oldrev}. 
The constraint (B.10) on $W$ via eq.~(B.9) is often taken
as the substitute to the generalized reality condition in eq.~(B.2) --- see, for example, eqs.
(A.19) and (A.22). However, it merely follows from eqs.~(B.2) and (B.9) that 
$\bo(\vec{L}-\vec{L}^{\dg})=0$ which generically implies that the harmonic function 
${\rm Im}\,\vec{L}$ is a constant,~\footnote{We assume that all the superfield components of
$W$ are regular in spacetime.}, ${\rm Im}\,\vec{L}=\vec{M}=const$. This constant $\vec{M}$ then
enters N=2 supersymmetry transformation laws, modifies the abelian constraints (A.19) as
$$ D^{ij}W-\bar{D}^{ij}\bar{W}=4iM^{ij}~,\eqno(B.12)$$ 
and it can be interpreted as a `{\it magnetic\/}' Fayet-Iliopoulos term \cite{apt}. It is not
clear to me, however, how the magnetic FI term could be introduced into the theory (3.8) since
the N=2 vector gauge superfield strength $W$ defined by eq.~(A.25) automatically satisfies
eq.~(B.12) with $M^{ij}=0$.

\end{document}